\documentclass[aps,prd,nofootinbib,reprint,superscriptaddress]{revtex4-1}
\usepackage{hyperref}
\usepackage{supertabular}
\usepackage{blindtext}
\usepackage{amsfonts}
\usepackage{tensor}
\usepackage{graphicx}
\usepackage{pict2e}
\usepackage{balance}

\usepackage{amssymb, amsmath,environ}

\usepackage{color}


\newcommand{\bse}{\begin{subequations}}
	\newcommand{\ese}{\end{subequations}}
\newcommand{\be}{\begin{equation}}
\newcommand{\ee}{\end{equation}}

\NewEnviron{eqsplit}{%
	\begin{equation}\begin{split}
	\BODY
	\end{split}\end{equation}
}

\makeatletter
\newcommand*\bigcdot{\mathpalette\bigcdot@{.5}}
\newcommand*\bigcdot@[2]{\mathbin{\vcenter{\hbox{\scalebox{#2}{$\m@th#1\bullet$}}}}}
\makeatother

\newcommand{\bea}{\begin{eqnarray}}
\newcommand{\eea}{\end{eqnarray}}
\newcommand{\ba}{\begin{array}}
	\newcommand{\ea}{\end{array}}

\newcommand{\la}{\langle}
\newcommand{\ra}{\rangle}

\begin{document}

	\title{Imprint of $\alpha$-Clustering on \textit{Ab Initio} Correlations in Relativistic Light Ion Collisions}
	
	\begin{abstract}
		This study investigates the influence of $\alpha$-cluster structures in relativistic light nuclear collisions. Using a cluster framework, I extract the characteristics of the nucleonic configurations of $^{16}$O and $^{20}$Ne as predicted by various \textit{ab-initio} models, including Nuclear Lattice Effective Field Theory (NLEFT), Variational Monte Carlo (VMC), and the Projected Generator Coordinate Method (PGCM). Additionally, I analyze configurations derived from a three-parameter Fermi (3pF) density function. The investigation focuses on the effects of cluster parameters on two-point correlators using a rotor model for symmetric collisions ($^{16}$O+$^{16}$O and $^{20}$Ne+$^{20}$Ne) and asymmetric collisions ($^{208}$Pb+$^{16}$O and $^{208}$Pb+$^{20}$Ne). The cluster parameters are determined by minimizing the \textit{chi-square} statistic to align the nucleon distributions with those predicted by the aforementioned theories. The results reveal that perturbative calculations effectively capture the structural features of these nuclei, while comparisons with Monte Carlo simulations validate these findings. Furthermore, the analysis reveals distinct cluster geometries: VMC suggests tetrahedral shapes, while NLEFT, PGCM, and 3pF indicate irregular triangular pyramids. Notably, NLEFT shows a bowling pin-like $\alpha$ cluster structure for $^{20}$Ne. The study also identifies constraints on cluster parameters in the different oxygen structures, with a gradual increase in $\varepsilon_2\{2\}$ for the states of $\alpha$+$^{12}$C. Accurate modeling of asymmetric collisions necessitates a range of nucleons from heavy spherical nuclei, leading to weighted correlators in perturbative calculations. I demonstrate consistency between perturbative calculations and Monte Carlo models, with analytical calculations providing more insights into asymmetric than symmetric collisions.	
	\end{abstract}

	\author{Hadi Mehrabpour}
	\email[]{hadi.mehrabpour.hm@gmail.com}
	\affiliation{School of Physics, Peking University, Beijing 100871, China}
	\affiliation{Center for High Energy Physics, Peking University, Beijing 100871, China}

	\maketitle

	\section{Introduction}\label{Introduction}
	Alpha clustering is a phenomenon predicted in the 1930s by George Gamow, which refers to the structural composition of nuclei made up of alpha-like four-nucleon clusters \cite{Gamow:1930}. The year 1937 marked a significant advancement in the understanding of alpha clustering, as Wilfried Wefelmeier observed that groups of nucleons form tightly bound sub-units that enhance stability and establish specific energy levels in light even-even nuclei, such as $^{16}$O and $^{20}$Ne \cite{Wefelmeier:1937,Nakanishi:1967,Ebran:2012ww}. These nuclei can be conceptualized as being formed by alpha clusters arranged in distinctive and regular geometries. This concept is particularly important for elucidating the structure of excited states \cite{Iachello:1981} and the formation of resonances in light nuclei \cite{Porro:2024bid}. Research in this domain frequently employs theoretical models, such as the cluster model \cite{Brink:1970,Tomoda:1978vlo,Descouvemont:1987zzb,Chernykh:2007zz,Kanada-Enyo:2019hrm}, which facilitate predictions regarding energy levels \cite{Bijker:2020}, decay processes \cite{Delion:2023qha}, and reaction mechanisms \cite{Herndl:1991zz} involving light nuclei. Furthermore, $\alpha$-clustering offers valuable insight into the intricate interactions within nuclei, thereby enhancing our understanding of nuclear structure and reactions in both astrophysics and nuclear physics \cite{PradaMoroni:2002um,Dominguez:2001kq,Imbriani:2001xz}. Notably, $^{16}$O and $^{20}$Ne are regarded as key nuclei in astrophysical contexts due to their clustering structures \cite{Herndl:1991zz,Lombardo:2023,deBoer:2017ldl,Lombardo:2019swy}.  
	
	Relativistic light nuclear collisions represent one of the most compelling methodologies in contemporary nuclear physics, facilitating the investigation of the structural characteristics inherent in colliding light nuclei, particularly those exhibiting $\alpha$-clustering \cite{Jia:2022ozr}. The initial-state geometry of the collisions is reflected in the final-state correlations of the emitted hadrons, thereby offering a unique probe into $\alpha$-cluster configurations of the light nuclei \cite{Broniowski:2013dia}. This has led to the study of light-ion structures having garnered significant attention in recent years \cite{Giacalone:2024luz,Duguet:2025hwi,Giacalone:2024ixe,Li:2025hae,Mehrabpour:2025rzt,Zhang:2024vkh,Shafi:2025feq,Lu:2025cni,R:2025qag}.  These studies focus on the effect of $\alpha$ clusters in $^{16}$ O and $^{20}$ Ne using \textit{ ab initio} models on flow observables. 
	Variations in Hamiltonians and approximations lead to significant differences in the structural predictions of light nuclei, as evidenced by calculations using NLEFT \cite{Meissner:2014lgi,Elhatisari:2017eno}, VMC \cite{Lonardoni:2018nob} and PGCM \cite{Frosini:2021ddm}. Consequently, these discrepancies result in divergent predictions for the observables of $^{16}$O and $^{20}$Ne in symmetric \cite{Zhang:2024vkh,Giacalone:2024luz} and asymmetric \cite{Giacalone:2024ixe} collisions.
	This study employs a cluster model introduced in Ref. \cite{Rybczynski:2017nrx} to demonstrate that these varying predictions stem from different parameters associated with $\alpha$-clustering, specifically the inter-cluster distance ($\ell_c$) and the size of the $\alpha$ particles ($r_L$), such that the ratio $r_L/\ell_c$ remains approximately constant. This study investigates the impact of these parameters on the initial correlators in head-on-head collisions involving $^{16}$O+$^{16}$O, $^{20}$Ne+$^{20}$Ne, $^{208}$Pb+$^{16}$O and $^{208}$Pb+$^{20}$Ne using three distinct approaches: perturbative calculations, a Monte Carlo simulation introduced in Ref. \cite{Giacalone:2023hwk}, and the TRENTo model as an initial-state generator \cite{Moreland:2014oya}. The details of the perturbative calculations will be outlined in Section \ref{Model setup}, where I will describe the methodology and provide the analytical forms of $k$-body density distributions and initial correlators related to the $\alpha$-cluster structure within a rigid rotor model. The results from both symmetric and asymmetric collisions will be presented in Sections \ref{sy} and \ref{asy}, respectively, followed by a summary of the findings in the concluding section.
	
	\begin{figure}[t!]
		\begin{tabular}{c}
			\includegraphics[scale=.8]{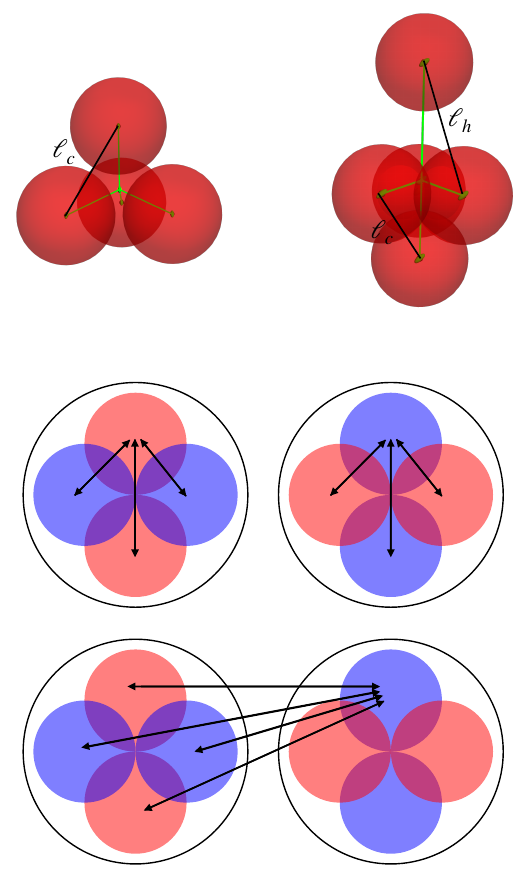}
		\end{tabular}
		\begin{picture}(0,0)
		\put(-190,40){{\fontsize{11}{11}\selectfont \textcolor{black}{Tetrahedron}}}
		\put(-70,40){{\fontsize{11}{11}\selectfont \textcolor{black}{Bowling-pin}}}
		\put(-225,160){{\fontsize{14}{14}\selectfont \textcolor{black}{(a)}}}
		\put(-225,20){{\fontsize{14}{14}\selectfont \textcolor{black}{(b)}}}
		\put(-225,-70){{\fontsize{11}{11}\selectfont \textcolor{black}{Intra-cluster}}}
		\put(-225,-170){{\fontsize{11}{11}\selectfont \textcolor{black}{Inter-cluster}}}
		\end{picture}		
		\caption{a) Schematic representations of the $\alpha$-cluster structures of $^{16}$O (tetrahedron) and $^{20}$Ne (bowling pin) are shown. Here, $\ell_c$ denotes the side length of the regular triangular pyramid in both structures, while $\ell_h$ indicates the distance from the center of the cluster at the top to the other three clusters in the middle of the bowling pin structure. Panel (b) illustrates two types of nucleon-nucleon correlations within a cluster model, intra-$\alpha$-cluster and inter-$\alpha$-cluster. The colors blue and red represent protons and neutrons, respectively. The top panel depicts the correlations between nucleons within the same clusters, whereas the bottom panel shows the correlations between nucleons from different clusters. These two effects contribute simultaneously to the results; however, for clarity, I distinguish them in this figure.} 
		\label{Fige1}
	\end{figure}
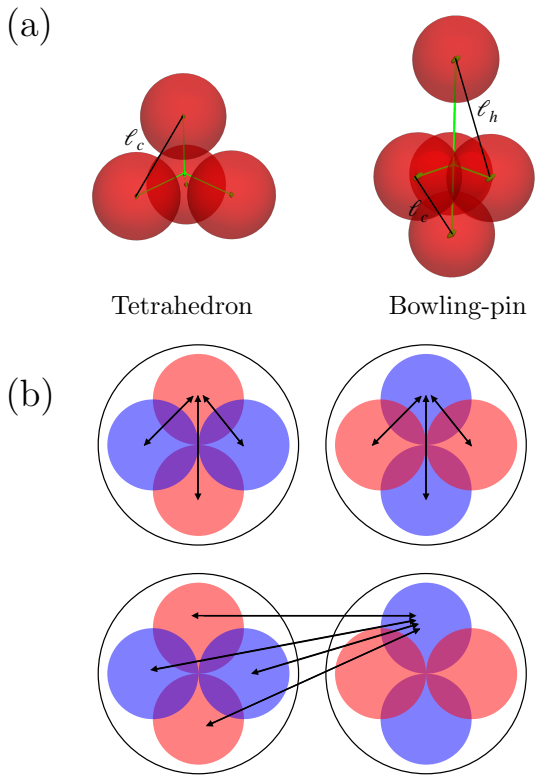

	\section{Methodology}\label{Model setup}
	Existence of $\alpha$-clustering in light ions has been strongly confirmed in low-energy nuclear experiments \cite{Youngblood:1997zz,Youngblood:1998zz,Lui:2001xh,Gupta:2015hpa,Yildiz:2006xc}. This phenomenon can be reproduced using modern \textit{ab-initio} models such as NLEFT, VMC, and PGCM \cite{YuanyuanWang:2024sgp}. Recent studies \cite{Giacalone:2024luz,Li:2025hae,Liu:2025zsi} in relativistic ion collisions indicate that these models predict a tetrahedron structure for $^{16}$O and a bowling-pin (BP) shape for $^{20}$Ne. Fig.\ref{Fige1}(a) shows a schematic of the $\alpha$ clusters in oxygen (left) and neon (right). I should mention that the parameters $\ell_c$ and $\ell_h$ generally indicate the distances between the centers of two clusters. In this way, this study aims to investigate the nucleon configurations generated by these models for $^{16}$O and $^{20}$Ne in the context of the cluster model introduced by Ref.\cite{Rybczynski:2017nrx}. 
	
	The encoded nuclear structure information in the collision zone can be reconstructed by employing hadron spectra from the detector. To achieve this, one should study the average transverse momenta of hadrons, defined as
	\begin{align}
	[p_T] = (1/N_{ch})\sum_{i=1}^{N_{ch}} p_{T,i},\label{Q1}
	\end{align}  
	where $N_{ch}$  is the total yield of hadrons in a collision. Additionally, the azimuthal distribution of the produced hadrons can be analyzed using a Fourier decomposition approach: 
	\begin{align}
	\frac{dN}{d\phi p_T dp_T}= \frac{dN}{p_T dp_T} \sum_{n=-\infty}^{\infty}V_n(p_T) e^{in\phi}, \quad  |V_n|=v_n.\label{Q2}
	\end{align} 
	In this context, the complex harmonics $V_n=v_n e^{in\psi_n}$, referred to as the coefficients of anisotropic flow, capture the anisotropic nature of particle emission. Here, $\psi_n$ denotes the event-plane angle associated with these harmonics. The hydrodynamic paradigm establishes a one-to-one relationship between initial- and final-state observables \cite{Niemi:2012aj,Giacalone:2020dln}. Consequently, it connects the fluctuations of $[p_T]$ and $V_n$ to the fluctuations of initial energy density $E$ and initial special anisotropies $\mathcal{E}_n$ of the overlap area in the transverse plane ($\mathbf{r}=(x,y)$):
	\begin{align}
	[p_T]&\propto E=\int_{\mathbf{r}} \epsilon(\mathbf{r}),\\
	V_n &\propto \mathcal{E}_n=-\frac{\int_{\mathbf{r}} |\mathbf{r}|^n e^{in\Phi_n}\epsilon(\mathbf{r})}{\int_{\mathbf{r}} |\mathbf{r}|^n\epsilon(\mathbf{r})},
	\end{align}
	where $|\mathbf{r}|=\sqrt{x^2+y^2}$ and $\Phi_n=\arctan2(y/x)$. Notice that $\epsilon(\mathbf{r})$ denotes a transverse energy density distribution in a single event. This leads to similar relations for moments,
	\begin{align}
	\la (\delta [p_T]/[p_T])^2\ra &\propto \text{var}(E/\la E\ra),\label{q03}\\
	\la V_n V_n^*\ra &\propto \la \mathcal{E}_n\mathcal{E}_n^*\ra.\label{q04} 
	\end{align} 
	Finding the observables in Eqs.\ref{q03} and \ref{q04} requires obtaining energy density in the collision zone. Lake of complication, in this study, I consider that $\epsilon(\mathbf{r}) = T(\mathbf{r}) T'(\mathbf{r})$ concerning the transverse profiles of the two colliding nuclei due to Lorentz-contraction, $T(\mathbf{r})$ and  $T'(\mathbf{r})$. A similar approach is also considered in the TRENTo model. To do this, I first find nucleon charge density to determine the positions of nucleons, and then I calculate the initial correlators of Eqs.\ref{q03} and \ref{q04} to study the $\alpha$ clusters in light nuclei \footnote{Notice that scrutinizing this type of nuclear structure may be required to consider higher order correlators \cite{Li:2025hae}. Here, I try to show a simple face of an analytical approach to study $\alpha$-clustering. Of course, one can extend this study concerning three-point correlators such as cov($\mathcal{E}_n\mathcal{E}_n^*, E$) \cite{Giacalone:2023hwk}.}.
	
	\subsection{Charge density}\label{char}
	To study light nuclear structures modeled by the $\alpha$-clustering approach, I consider the distribution of the nucleons in each cluster according to the Gaussian function \cite{Rybczynski:2017nrx}:
	\begin{equation}\label{q1}
	\rho_{\alpha_i}(\vec{r}) = \Big(\frac{3}{2\pi r_L^2}\Big)^{3/2} \exp\Big[-\frac{3(\vec{r}-\vec{L}_i)^2}{2r_L^2}\Big],
	\end{equation}
	where $\vec{r}$ is the 3D coordinate of the nucleon, $\vec{L}_i$ is the position of the center of the cluster $i$, and $r_L$ is the root-mean-square radius of the cluster. To have the charge density of light nuclei in the context of $\alpha$-clustering, it is required $\rho_N(\vec{r})=\sum_i^{n} \rho_{\alpha_i}(\vec{r})$ for $n$ clusters. Since nuclei generally have an arbitrary orientation in a given event, we can construct this phenomenon using Euler angles in our calculations. There are six possibilities for choosing the rotation of axes for the three Euler angles, where the first and third rotation axes are the same. In this work, I choose Z-X-Z rotation, $\Omega=(a_1,a_2,a_3)$, such that I rotate nuclei in this scheme. The transverse $k$-body density of nucleons can be found by integrating out the z component and the Euler angles (or orientational averaging) \cite{Alvioli:2009ab}:
	\begin{equation} \label{q2}
	\begin{split} 
	&\rho_{\perp}^{(k)}(\mathbf{r}_1,\cdots,\mathbf{r}_k) =\\ &\int dz_1\;\cdots\int dz_k\; \int d\Omega\; \rho_N(R_{zxz}(\Omega)\vec{r_1})\cdots\rho_N(R_{zxz}(\Omega)\vec{r_k}),
	\end{split} 
	\end{equation}
	where $R_{zxz}$ is the related rotational matrix. Since I here study only the effect of $\alpha$-clusters on the two-particle correlation functions, I need to find one- and two-body densities. In the following, I explain how to obtain these densities and what challenges exist there, separately.  
	\\\textbf{\textit{One-body density:}} The transverse one nucleon distribution inside a light nuclear modeled by $\alpha$-clustering approach is defined as follows:
	\begin{equation}\label{rhoT1}
	\begin{split}
	\rho_{\perp}^{(1)}(\mathbf{r},\Omega)&=\int dz\; \rho_N(\vec{r},\Omega)\\&=\sum_{i=1}^{N_{\alpha}}\int dz\; \rho_{\alpha_i}(\vec{r},\Omega)\\&=\sum_{i=1}^{N_{\alpha}}\rho_{\perp}^{(\alpha_i)}(\mathbf{r},\Omega),
	\end{split}	
	\end{equation}
	where $N_{\alpha}$ displays the number of clusters	\footnote{Since the averaging over the Euler angle is not doable exactly, I leave it for a numerical integration in the final step after finding observables.}. This will also be considered for two-body correlations. To find the transverse charge density Eq.\ref{rhoT1}, I rotate the charge density of Eq.\ref{q1} using the rotational matrix $R_{zxz}(\Omega)$ and then I integrate out the z-component. Thus, one can find:
	\begin{equation}\label{q3}
	\begin{split}
	&\rho_{\perp}^{(\alpha_i)}(\mathbf{r})=\la\rho_{\perp}^{(\alpha_i)}(\mathbf{r},\Omega) \ra_{\Omega}=\\ &\la\frac{3}{2\pi r_L^2}\exp\Big[-\frac{3(x^2+y^2+f_i-2(x h_{x,i}+y h_{y,i}))}{2 r_L^2}\Big]\ra_{\Omega},
	\end{split}
	\end{equation}  
	where $\la\cdots\ra_{\Omega}$ is the average over the Euler angles. The depicted functions $f_i$, $h_{x,i}$, and $h_{y,i}$ in Eq.\ref{q3} are:
	\begin{align}
	f_i &= |\vec{L}_i|^2-(L_{z,i}c2+(-L_{y,i}c1+L_{x,i}s1)s2)^2,\label{fi}\\
	\begin{split}
	h_{x,i}&= (L_{y,i}c1c2-L_{x,i}c2s1+L_{z,i}s2)s3\\&+(L_{y,i}s1+L_{x,i}c1)c3,\label{hx}
	\end{split}\\
	\begin{split} 
	h_{y,i}&= (L_{y,i}c1c2-L_{x,i}c2s1+L_{z,i}s2)c3\\&-(L_{y,i}s1+L_{x,i}c1)s3,\label{hy}
	\end{split}
	\end{align}
	with $ci=\cos(a_i)$ and $si=\sin(a_i)$. We note that the positions of cluster centers are found regarding their distances from the origin, $|\vec{L}_i|$ (see Table \ref{tab1}). Having Eq.\ref{q3} one can study the effects of cluster parameters on the one-body transverse distribution of nucleons. 
	\\\textit{\textbf{Two-body density:}} Finding two-particle correlators is required to obtain the transverse 2-body density of nucleons,
	\begin{equation}\label{q4}
	\begin{split}
	\rho_{\perp}^{(2)}(\mathbf{r},\mathbf{r}')&=\int d\Omega\int dz\int dz'\;\rho_N(R_{zxz}(\Omega)\vec{r})\rho_N(R_{zxz}(\Omega)\vec{r'})\\
	&=\sum_{i,j}^{N_{\alpha}}\la \rho_{\perp}^{(\alpha_i)}(\mathbf{r},\Omega)\rho_{\perp}^{(\alpha_j)}(\mathbf{r}',\Omega)\ra_{\Omega}\\
	&=\frac{9}{4\pi^2  r_L^4}e^{-\frac{3(|\mathbf{r}|^2+|\mathbf{r}'|^2)}{2 r_L^2}}\\&\hspace*{1.5cm}\times\sum_{i,j}^{N_{\alpha}}\la e^{\frac{f_i+x h_{x,i}+y h_{y,i}+f_j+x' h_{x',j}+y' h_{y',j}}{r_L^2/3}}\ra_{\Omega}.
	\end{split} 
	\end{equation}
	In the study of light nuclear structures governed by $\alpha$-clusters, we should consider two different correlations to cover all elements of two-point functions (see also Appendix \ref{app1}): 
	\begin{align}
	\rho_{\perp,1}^{(2)}(\mathbf{r},\mathbf{r}')&=\sum_{i=1}^{N_{\alpha}}\la \rho_{\perp}^{(\alpha_i)}(\mathbf{r},\Omega)\rho_{\perp}^{(\alpha_i)}(\mathbf{r}',\Omega)\ra_{\Omega},\label{q5}\\
	\rho_{\perp,2}^{(2)}(\mathbf{r},\mathbf{r}')&=\sum_{i\neq j}^{N_{\alpha}}\la \rho_{\perp}^{(\alpha_i)}(\mathbf{r},\Omega)\rho_{\perp}^{(\alpha_j)}(\mathbf{r}',\Omega)\ra_{\Omega}.\label{q6}
	\end{align} 
	Eq.\ref{q5} indicates the correlations between the nucleons that are inside a given $\alpha$-cluster (intra-$\alpha$), top panel in Fig.\ref{Fige1}(b),  such that this corresponds to $N_{\alpha}$ terms. $\rho_{\perp,2}^{(2)}(\mathbf{r},\mathbf{r}')$ in Eq.\ref{q6} says about the correlations between the nucleons located in different clusters (inter-$\alpha$), as depicted in the bottom panel of Fig.\ref{Fige1}(b). This type contains $N_{\alpha}(N_{\alpha}-1)$ terms. In other words, it is calculated concerning the off-diagonal elements of the correlations where $i\neq j$ in Eq.\ref{q4}. This categorization indicates that the study of $k$-body correlations (for $k\geq2$) for $\alpha$-clusters in light nuclei is involved in a different approach from which is presented in Ref.\cite{Giacalone:2023hwk}. I demonstrate this difference when I compare the analytical calculations with Monte Carlo simulations.   
	\\\textit{\textbf{Spherical shape:}} To study the case of spherical nuclei, I assume a cluster included more than 4 nucleons (e.g., 16 nucleons for oxygen) such that the radius of this cluster is determined by a corrected radius $R_s$ \cite{Duguet:2025hwi},
	\begin{equation}\label{Qsph} 
	N(x,y,z)=\Big(\frac{3}{2\pi R_s^2}\Big)^{3/2} \exp\Big[-\frac{3(x^2+y^2+z^2)}{2R_s^2}\Big].
	\end{equation}
	In Sec.\ref{res}, I will explain how $R_s$ is obtained
	once I work with a 3-parameter Fermi (3pF) density distribution: 
	\begin{equation}\label{3pf}
	N_{\text{3pF}}(r) \propto \frac{1+w(r^2/R_0^2)}{1+e^{(r-R_0)/a_0}},
	\end{equation}
	where $R_0$ denotes the half-width radius, $w$ measures central density depletion, and $a_0$ characterizes the surface diffuseness. The density $N(x,y,z)$ in Eq.\ref{Qsph} is invariant under the rotations $R_{zxz}(\Omega)$, therefore the $k$-body densities in the transverse plane are defined as a multiple of one-body densities: 
	\begin{equation}\label{Q8} 
	\begin{split}
	N_{\perp}^{(k)}(\mathbf{r}_1,\cdots,\mathbf{r}_k)&=\int dz_1\cdots dz_k \int d\Omega\; N(\vec{r_1})\cdots N(\vec{r_k})\\
	&= \Big(\frac{3}{2\pi R_s^2}\Big)^k\prod_{i}^{k} e^{-\frac{|\mathbf{r}_i|^2}{2R_s^2/3}}.
	\end{split} 
	\end{equation}
	This also provides a potential to study asymmetric collisions such as $^{208}$Pb+$^{16}$O and $^{208}$Pb+$^{20}$Ne, which have collided spherical ions. These collisions provide a chance to investigate light nuclear structures in high multiplicity or equivalently high resolution at SMOG2 experiment \cite{Jia:2022ozr}. Therefore, I will also study this type of collision in Sec.\ref{res}. Before going to discuss results, first, I present the form of initial correlators Eqs.\ref{q03} and \ref{q04} in the next part using calculated one- and two-body densities in the transverse plane.   
	
	\subsection{Correlators}\label{corr}
	In this section, I study the effect of cluster parameters on the fluctuations of $E$ and $\mathcal{E}_2$. To do so, the energy density in a given event class is written as \cite{Blaizot:2014nia}:
	\begin{equation}\label{energyF}
	\epsilon(\mathbf{r})= \la \epsilon(\mathbf{r})\ra +\delta \epsilon(\mathbf{r}),
	\end{equation}
	where $\la \epsilon(\mathbf{r})\ra$ is the average energy density and $\delta\epsilon(\mathbf{r})$ denotes the fluctuations. Having Eq.\ref{energyF}, I calculate the initial correlators $\la E\ra_{ev}$, Var($E$) and $\la \mathcal{E}_2\mathcal{E}_2^*\ra_{ev}$ using Eqs.\ref{rhoT1}, \ref{q5}, \ref{q6}, and \ref{Q8} mentioned in Sec.\ref{char}. The event average of the total system's energy at zero impact parameter can be obtained concerning the local average of the energy density in a given event sample \footnote{Of course, one can consider the effect of impact parameter and study these calculations for other centralities.}\footnote{We note that $\la \delta\epsilon(\mathbf{r})\ra_{ev}=0$ \cite{Blaizot:2014nia,Giacalone:2023hwk}.}: 
	\begin{equation}\label{A2}
	\begin{split}
	\la E\ra_{ev}&=\int d\Omega\;d\Omega'\;\la E(\Omega,\Omega')\ra\\&=\int d\Omega\;d\Omega'\;\la\epsilon(\mathbf{r},\Omega,\Omega')\ra
	\\&=\int d\Omega\;d\Omega'\;\la\sum_{s=1}^{A}\sum_{t=1}^{B}\mathcal{G}(\mathbf{r}-\xi_s)\mathcal{G}(\mathbf{r}-\xi_t)\ra,
	\end{split}	
	\end{equation}
	where $\mathcal{G}(\mathbf{r}-\xi_s)\equiv\frac{1}{2\pi w^2} \exp\left(-\frac{(x-x_s)^2+(y-y_s)^2}{2w^2}\right)$ is the standard two-dimensional high-energy gluonic form factor \cite{Moreland:2014oya} with a nucleon size $w$. $A\equiv N_{part,P}$ and $B\equiv N_{part,T}$ are the number of participants from projectile and target nuclei in the collision area. Since I do not integrate out the Euler angles until the end, first I calculate one-point function $\la\epsilon(\mathbf{r})\ra$ as a function of $\Omega$: 
	\begin{equation}\label{A3}
	\begin{split}
	&\la\epsilon(\mathbf{r},\Omega,\Omega')\ra\\&=AB\int d^2\xi_s\;d^2\xi_t\;\mathcal{G}(\mathbf{r}-\xi_s)\rho_{\perp}^{(1)}(\xi_s,\Omega) \mathcal{G}(\mathbf{r}-\xi_t)\rho_{\perp}^{(1)}(\xi_t,\Omega')\\
	&=AB\sum_{i,j=1}^{N_{\alpha}}\int d^2\xi_s\;d^2\xi_t\
	\\&\hspace*{2cm}\times\mathcal{G}(r-\xi_s)\rho_{\perp}^{(\alpha_i)}(\xi_s,\Omega) \mathcal{G}(r-\xi_t)\rho_{\perp}^{(\alpha_j)}(\xi_t,\Omega')\\
	&=AB\sum_{i,j=1}^{N_{\alpha}}I_i(\mathbf{r},\Omega)I_j(\mathbf{r},\Omega').
	\end{split}
	\end{equation}
where
\begin{equation}
	\begin{split}
	I_i(\mathbf{r},\Omega) &= \int \;d^2\xi_t\mathcal{G}(r-\xi_s)\rho_{\perp}^{(\alpha_i)}(\xi_s,\Omega)
	\\&= \frac{3}{2 \pi  \left(r_L^2+3 w^2\right)}e^{-\frac{3 (x^2+y^2)}{2 \left(r_L^2+3 w^2\right)}}e^{\frac{3 (xh_{x,i}+yh_{y,i})}{r_L^2+3 w^2}}\\&\hspace*{2cm}\times e^{-\frac{3 \left(f_i \left(r_L^2+3 w^2\right)-3 h_{x,i}^2 w^2-3 h_{y,i}^2 w^2\right)}{2 r_L^2 \left(r_L^2+3 w^2\right)}}.
	\end{split} 
\end{equation}
Notice that in Eq.\ref{A3}, I supposed all the nucleons from the projectile and target participated in the collisions. This assumption is valid only once we are working with symmetric collisions at zero impact parameters. Thus, by replacing Eq.\ref{A3} in Eq.\ref{A2}, we have: 
	\begin{equation}\label{A4}
	\begin{split}
	\la E\ra_{ev}&=AB\int d\Omega\;d\Omega'\;d^2\mathbf{r}\;\la\epsilon(\mathbf{r},\Omega,\Omega')\ra\\&=AB\sum_{i,j=1}^{N_{\alpha}}\int d\Omega\;d\Omega'\;d^2\mathbf{r}\;I_i(\mathbf{r},\Omega)I_j(\mathbf{r},\Omega').
	\end{split} 
	\end{equation}
	Concerning the coefficients in Eqs.\ref{fi}-\ref{hy}, the general form of $\la E\ra_{ev}$ is obtained as follows:
	\begin{equation}\label{q20}
	\begin{split}
	\la E\ra_{ev}&=\frac{3AB}{2\pi(r_L^2+r_L'^2+6w^2)}
	\\&\hspace*{1.5cm}\times\sum_{i,j=1}^{N_{\alpha}}\int d\Omega\;d\Omega'\ e^{f_i+f_j+\frac{F(\Omega,\Omega',\vec{L}_i,\vec{L}_j)}{6(r_L^2+r_L'^2+6w^2)}},
	\end{split} 
	\end{equation}
	where the function $F$ is:
	\begin{equation*}
	\begin{split}
	F(\Omega,\Omega',\vec{L}_i,\vec{L}_j)&=r_L^2r_L'^2 ((h_{x,i}+h_{x,j})^2+(h_{y,i}+h_{y,j})^2)\\
	&+6w^2((h_{x,i}+h_{y,i})^2r_L^2+(h_{x,j}+h_{y,j})^2r_L'^2).
	\end{split}
	\end{equation*}
	As an example, Eq.\ref{q20} can be found for the spherical ion collisions with the half-density radius $R_s$:
	\begin{equation}\label{q25} 
	\la E\ra_{ev} = \frac{3A_s^2}{4\pi (R_s^2+3 w^2)}. 
	\end{equation}
	This reveals that the average energy density has an inverse relation with $R_s$ and nucleon width $w$. We note that $A_s$ refers to the total number of nucleons within the nuclei that participate in the collisions. This is different from what we have in Eq. \ref{q20}, where  A  and  B  indicate the number of nucleons within a cluster. For more details about the relationship between the cluster parameters and the spherical parameters, please refer to Appendix \ref{app2}.  
	Finding two-particle correlators is required to evaluate the connected two-point functions of the field:
	\begin{equation}\label{A7}
	\begin{split}
	&\la\epsilon(\mathbf{r},\Omega,\Omega')\epsilon(\mathbf{r}',\Omega,\Omega')\ra
	\\&=\la\sum_{s,t=1}^{A,B}\mathcal{G}(\mathbf{r}-\xi_s)\mathcal{G}(\mathbf{r}-\xi_t)\sum_{n,m=1}^{A,B}\mathcal{G}(\mathbf{r}'-\xi_n)\mathcal{G}(\mathbf{r}'-\xi_m)\ra
	\\&=\la\sum_{s,n=1}^{A}\mathcal{G}(\mathbf{r}-\xi_s)\mathcal{G}(\mathbf{r}'-\xi_n)\sum_{t,m=1}^{B}\mathcal{G}(\mathbf{r}-\xi_t)\mathcal{G}(\mathbf{r}'-\xi_m)\ra
	\\&=F_P F_T,
	\end{split}
	\end{equation}
	where $F_X=N_{part,X}\;\mathcal{I}_{1,X}(\mathbf{r},\mathbf{r}',\Omega)+(N_{part,X}^2-N_{part,X})\;\mathcal{I}_{2,X}^{(1)}(\mathbf{r},\mathbf{r}',\Omega))+N_{part,X}^2\;\mathcal{I}_{2,X}^{(2)}(\mathbf{r},\mathbf{r}',\Omega)$. The indexes $P$ and $T$ denote projectile and target nuclei, respectively. Notice that in this case against large nuclei (see Eq.37 in Ref.\cite{Giacalone:2023hwk}), we must consider two different correlations, encoded in $\mathcal{I}_{2,X}^{(1)}$ and $\mathcal{I}_{2,X}^{(2)}$, as I mentioned in Eqs.\ref{q5} and \ref{q6} and depicted in Fig.\ref{Fige1}(b). Since $\la\epsilon(\mathbf{r},\Omega,\Omega')\epsilon(\mathbf{r}',\Omega,\Omega')\ra$ involves the transverse one- and two-body densities, we have:  
	\begin{align}
	\begin{split}
	\mathcal{I}_{1,X}&= \int d^2\xi_s\;\rho_{\perp}^{(1)}(\xi_s,\Omega)\mathcal{G}(\mathbf{r}-\xi_s) \mathcal{G}(\mathbf{r}'-\xi_s)\\
	&=\sum_{i=1}^{N_\alpha}\int d^2\xi_s\;\rho_{\perp}^{\alpha_i}(\xi_s,\Omega)\mathcal{G}(\mathbf{r}-\xi_s) \mathcal{G}(\mathbf{r}'-\xi_s)\\
	&=\sum_{i=1}^{N_\alpha}H_i^{(X)}(\mathbf{r},\mathbf{r}',\Omega),
	\end{split}
	\end{align}
	\begin{align} 
	\begin{split} 
	\mathcal{I}_{2,X}^{(1)} &= \int d^2\xi_s\;d^2\xi_n\;\rho_{\perp,1}^{(2)}(\xi_s,\xi_n,\Omega)\mathcal{G}(\mathbf{r}-\xi_s) \mathcal{G}(\mathbf{r}'-\xi_n)\\
	&=\sum_{i=1}^{N_\alpha}\int d^2\xi_s\;d^2\xi_n\;\rho_{\perp}^{(\alpha_i)}(\xi_s,\Omega)\rho_{\perp}^{(\alpha_i)}(\xi_n,\Omega)\\&\hspace*{3cm}\times\mathcal{G}(\mathbf{r}-\xi_s) \mathcal{G}(\mathbf{r}'-\xi_s)\\
	&=\sum_{i=1}^{N_\alpha}H_{ii}^{(X)}(\mathbf{r},\mathbf{r}',\Omega),
	\end{split}\\
	\begin{split} 
	\mathcal{I}_{2,X}^{(2)} &= \int d^2\xi_s\;d^2\xi_n\;\rho_{\perp,2}^{(2)}(\xi_s,\xi_n,\Omega)\mathcal{G}(\mathbf{r}-\xi_s) \mathcal{G}(\mathbf{r}'-\xi_n)\\
	&=\sum_{i\neq j=1}^{N_\alpha}\int d^2\xi_s\;d^2\xi_n\;\rho_{\perp}^{(\alpha_i)}(\xi_s,\Omega)\rho_{\perp}^{(\alpha_j)}(\xi_n,\Omega)\\&\hspace*{3cm}\times\mathcal{G}(\mathbf{r}-\xi_s) \mathcal{G}(\mathbf{r}'-\xi_s)\\
	&=\sum_{i\neq j=1}^{N_\alpha}H_{ij}^{(X)}(\mathbf{r},\mathbf{r}',\Omega).
	\end{split} 
	\end{align}
	We note that $\mathcal{I}_1$, $\mathcal{I}_2^{(1)}$, and $\mathcal{I}_2^{(2)}$ indicate the effect of one-body density, two-body density within a cluster, and two-body correlations between the nucleons of various clusters, respectively. By assuming whole nucleons participated in the collisions and replacing them in Eq.\ref{A7}, the two-point function $\la\varepsilon(\mathbf{r})\varepsilon(\mathbf{r}')\ra$ is rewritten as follows:
	\begin{equation}\label{A8}
	\begin{split}
	&\la\epsilon(\mathbf{r},\Omega,\Omega')\epsilon(\mathbf{r}',\Omega,\Omega')\ra
	\\&=AB\sum_{i,i'}^{N_\alpha}H_i^{(P)}(\mathbf{r},\mathbf{r}',\Omega)H_{i'}^{(T)}(\mathbf{r},\mathbf{r}',\Omega')\\&+2A(B^2-B)\sum_{i,i'}^{N_\alpha}H_i^{(P)}(\mathbf{r},\mathbf{r}',\Omega)H_{i'i'}^{(T)}(\mathbf{r},\mathbf{r}',\Omega')\\&+2AB^2\sum_{i,i'\neq j'}^{N_\alpha}H_i^{(P)}(\mathbf{r},\mathbf{r}',\Omega)H_{i'j'}^{(T)}(\mathbf{r},\mathbf{r}',\Omega')\\&+2A^2(B^2-B)\sum_{i\neq j,i'}^{N_\alpha}H_{ij}^{(P)}(\mathbf{r},\mathbf{r}',\Omega)H_{i'i'}^{(T)}(\mathbf{r},\mathbf{r}',\Omega')\\&+A^2B^2\sum_{i\neq j,i'\neq j'}^{N_\alpha}H_{ij}^{(P)}(\mathbf{r},\mathbf{r}',\Omega)H_{i'j'}^{(T)}(\mathbf{r},\mathbf{r}',\Omega')\\&+(A^2-A)(B^2-B)\;\sum_{i,i'}^{N_\alpha}H_{ii}^{(P)}(\mathbf{r},\mathbf{r}',\Omega)H_{i'i'}^{(T)}(\mathbf{r},\mathbf{r}',\Omega').
	\end{split}
	\end{equation} 
	As mentioned in the previous part, the other aim of this work is the study of asymmetric collisions, Pb+O and Pb+Ne. In this type of collision, one of the collided nuclei is spherical as the projectile, therefore the related term $F_P$ in Eq.\ref{A7} would be changed as $F_P=A\;\mathcal{I}_{1,P}(\mathbf{r},\mathbf{r}')+(A^2-A)\;\mathcal{I}_{2,P}(\mathbf{r},\mathbf{r}'))$ such that we have: 
	\begin{align*}
	\mathcal{I}_{1,P}&= \int d^2\xi_s\;N_{\perp}^{(1)}(\xi_s)\mathcal{G}(\mathbf{r}-\xi_s) \mathcal{G}(\mathbf{r}'-\xi_s)\\
	& = \frac{3}{4\pi^2w^2(R_s^2+3w^2)^2}e^{-\frac{R_s^2((x-x')^2+(y-y')^2)+3w^2(|\mathbf{r}|^2+|\mathbf{r}'|^2)}{2(2R_s^2w^2+3w^4)}},\\
	\mathcal{I}_{2,P} &= \int d^2\xi_s\;d^2\xi_n\;N_{\perp}^{(2)}(\xi_s,\xi_n)\mathcal{G}(\mathbf{r}-\xi_s) \mathcal{G}(\mathbf{r}'-\xi_n)\\
	&= \frac{9}{4\pi^2(R_s^2+3w^2)^2}\exp\Big[-\frac{3(|\mathbf{r}|^2+|\mathbf{r}'|^2)}{2(R_s^2w^2+3w^4)}\Big].
	\end{align*}
	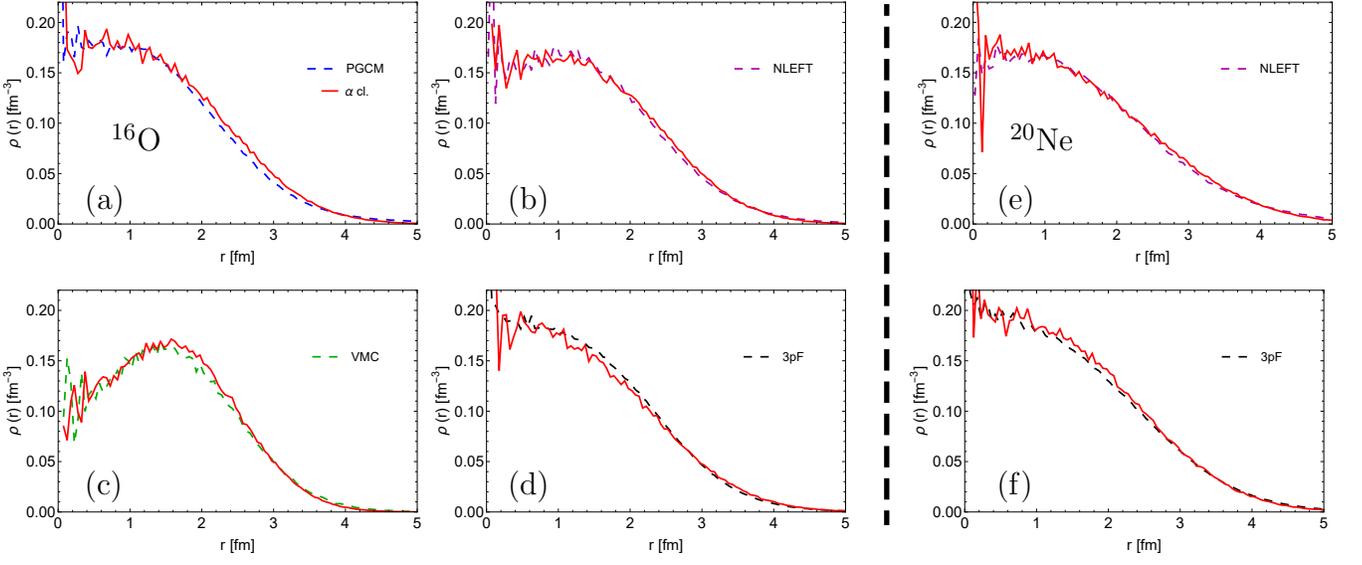
\begin{figure*}[t!]
		\begin{tabular}{c}
			\includegraphics[scale=.49]{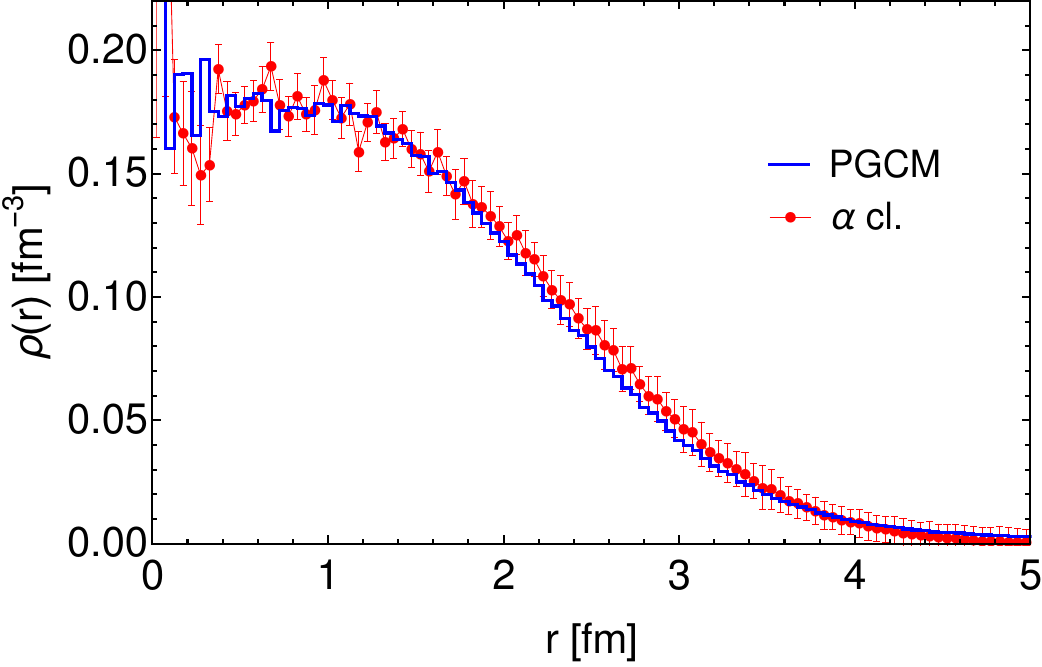}
			\includegraphics[scale=.49]{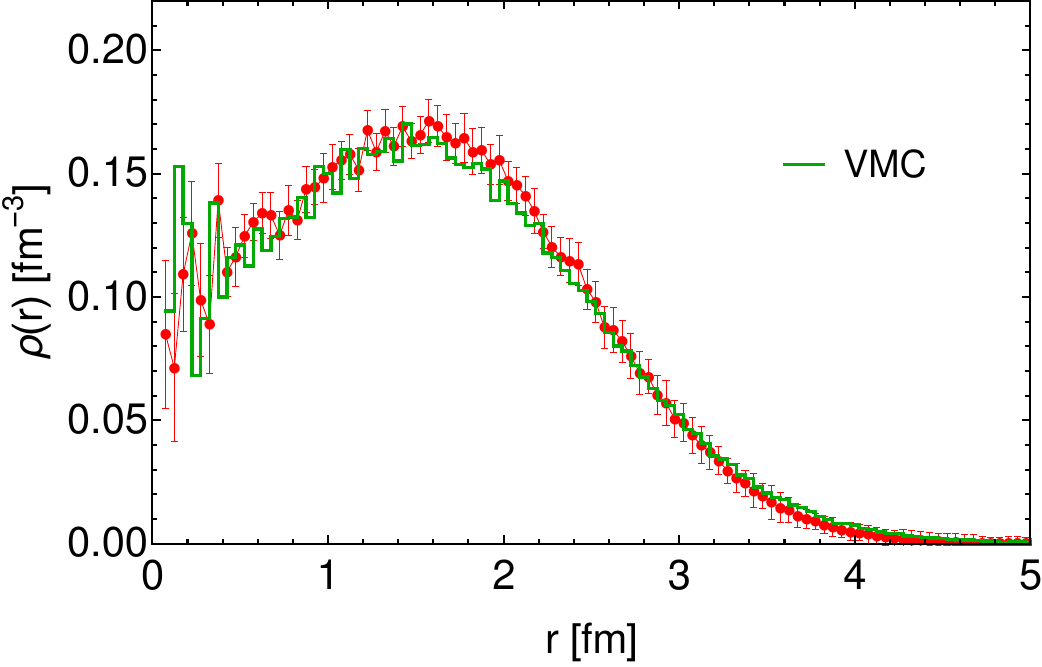}\\
			\includegraphics[scale=.49]{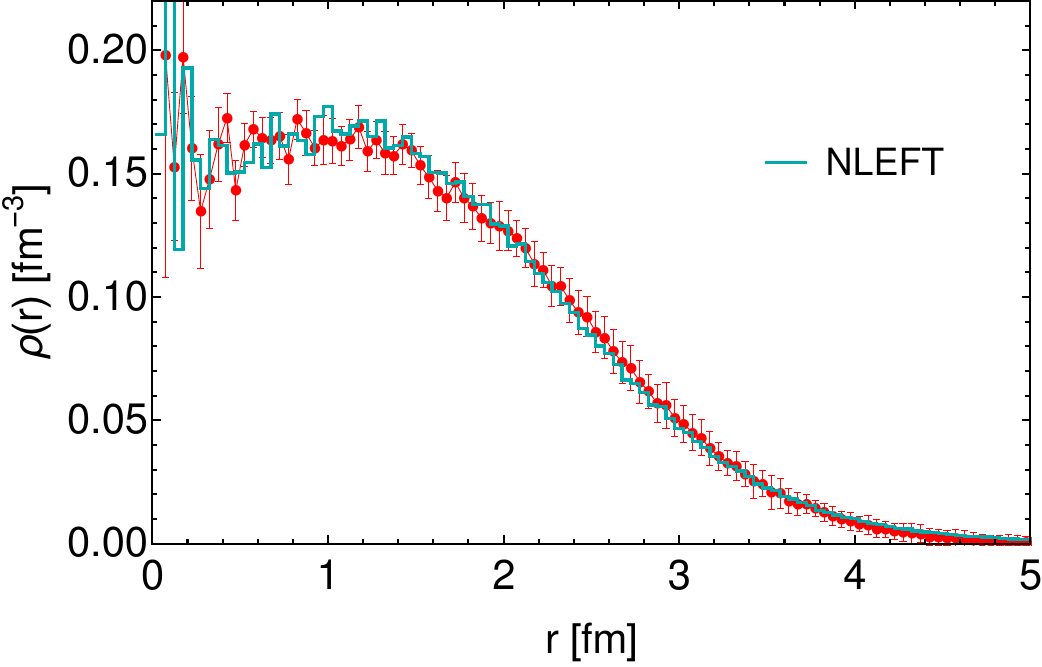}
			\includegraphics[scale=.49]{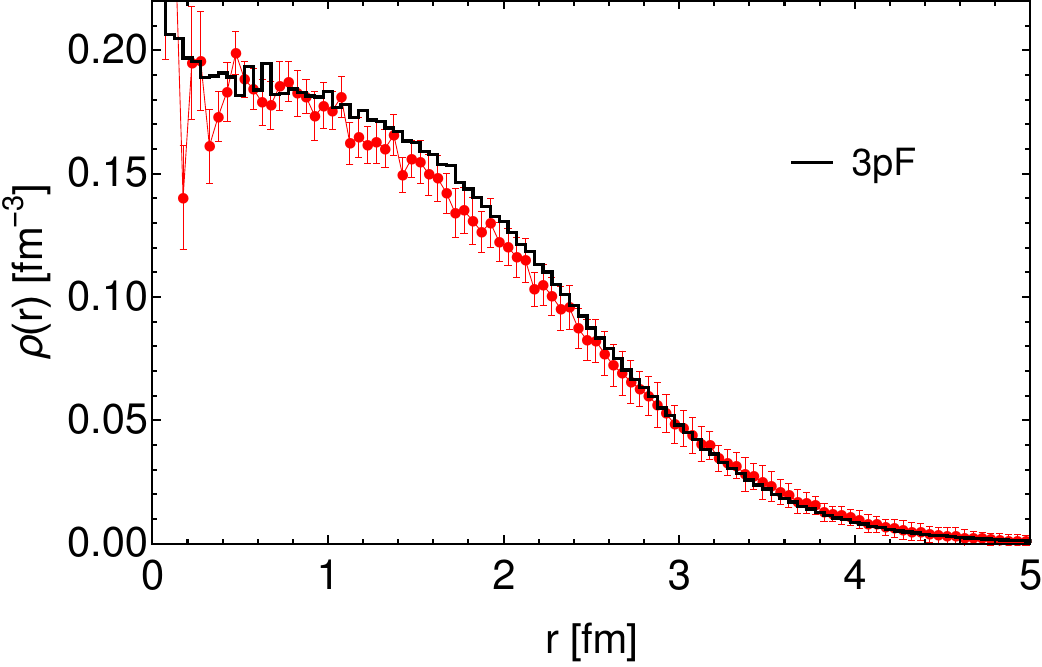}\\
			\includegraphics[scale=.49]{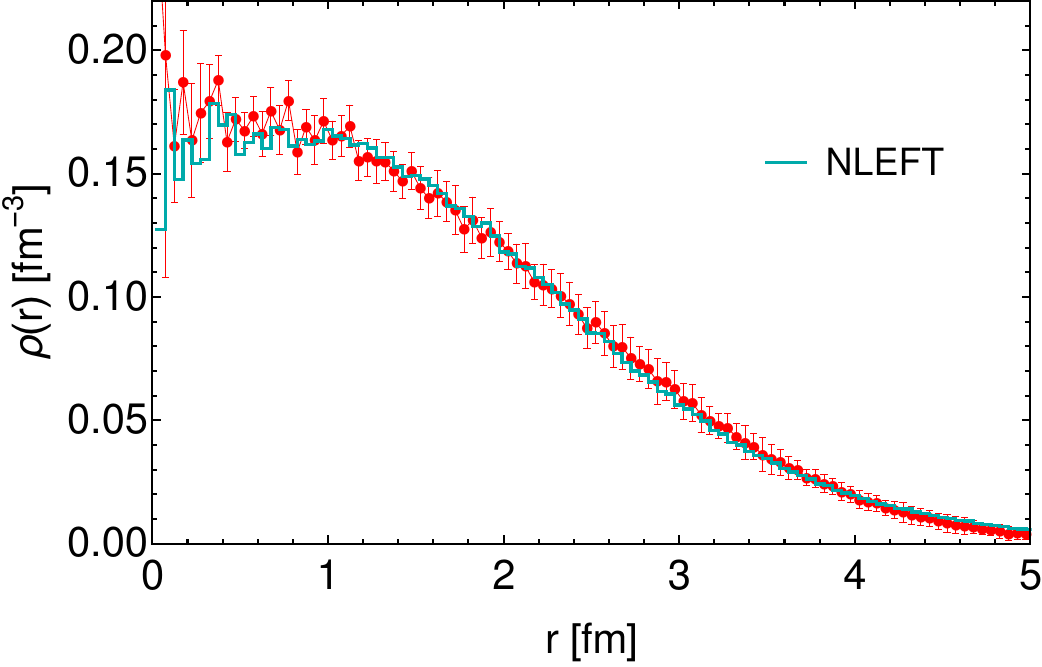}
			\includegraphics[scale=.49]{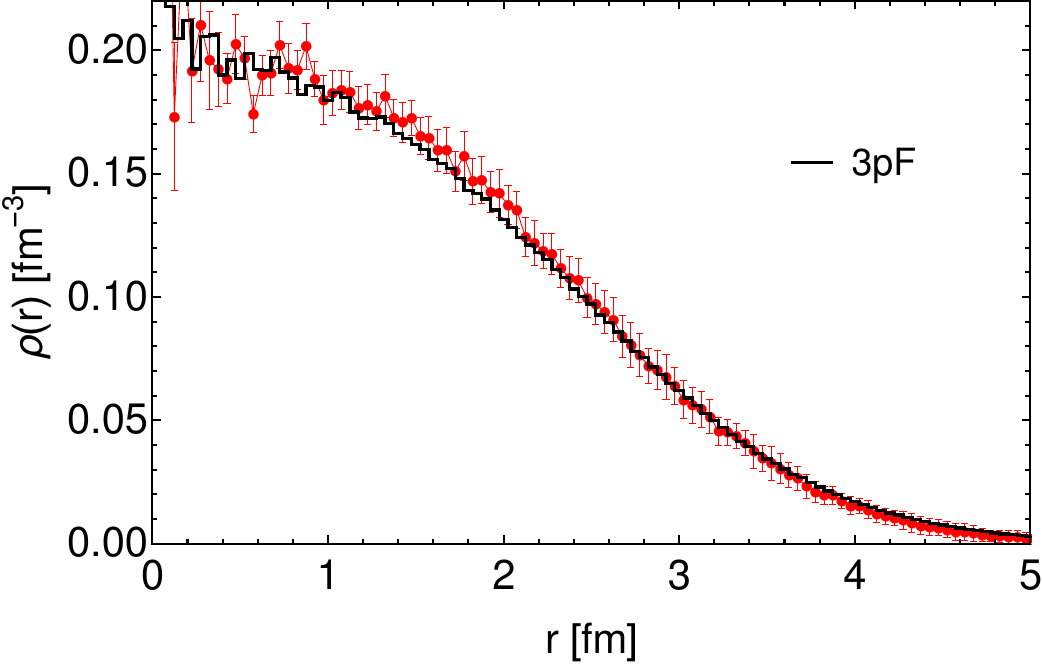}
		\end{tabular}
		\begin{picture}(0,0)
		\put(-453,123){{\fontsize{14}{14}\selectfont \textcolor{black}{(a)}}}
		\put(-203,123){{\fontsize{14}{14}\selectfont \textcolor{black}{(b)}}}
		\put(-453,-37){{\fontsize{14}{14}\selectfont \textcolor{black}{(c)}}}
		\put(-203,-37){{\fontsize{14}{14}\selectfont \textcolor{black}{(d)}}}
		\put(-453,-198){{\fontsize{14}{14}\selectfont \textcolor{black}{(e)}}}
		\put(-203,-198){{\fontsize{14}{14}\selectfont \textcolor{black}{(f)}}}
		\put(-350,220){{\fontsize{13}{13}\selectfont \textcolor{black}{$^{16}$O}}}
		\put(-100,220){{\fontsize{13}{13}\selectfont \textcolor{black}{$^{16}$O}}}
		\put(-350,60){{\fontsize{13}{13}\selectfont \textcolor{black}{$^{16}$O}}}
		\put(-100,60){{\fontsize{13}{13}\selectfont \textcolor{black}{$^{16}$O}}}
		\put(-350,-100){{\fontsize{13}{13}\selectfont \textcolor{black}{$^{20}$Ne}}}
		\put(-100,-100){{\fontsize{13}{13}\selectfont \textcolor{black}{$^{20}$Ne}}}
		\end{picture}		
		\caption{The one-nucleon density distributions are illustrated for $^{16}O$ (using PGCM in panel (a), VMC (b), NLEFT (c), 3pF (d)) and $^{20}Ne$ (using NLEFT (e), 3pF (f)). The red points represent the nuclear radial densities derived from the cluster configurations (labeled by $\alpha$ cl.). The curves for 3pF and PGCM are obtained from sampled configurations that have been re-centered to ensure an apples-to-apples comparison with the NLEFT and VMC results.} 
		\label{Fige2}  
	\end{figure*}
	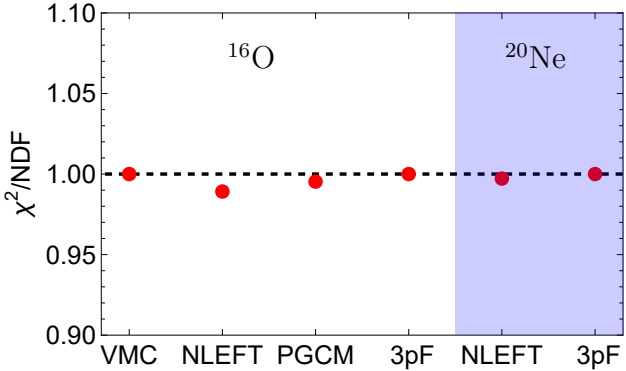
\begin{figure}[t!]
		\begin{tabular}{c}
			\includegraphics[scale=.65]{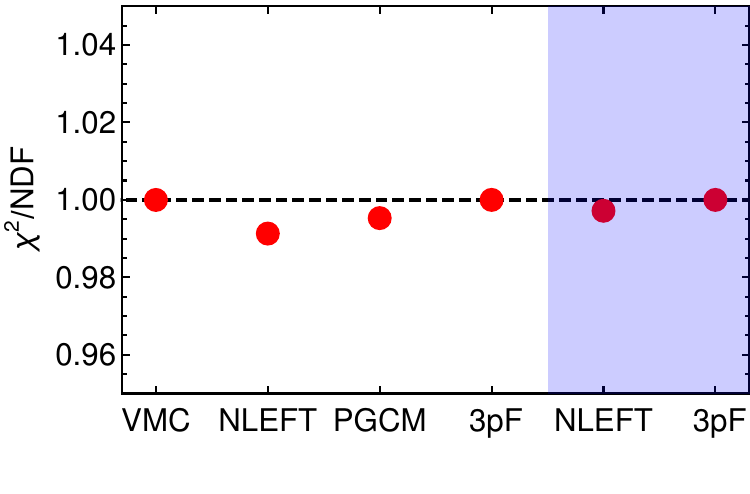}
		\end{tabular}
		\begin{picture}(0,0)
		\put(-155,55){{\fontsize{12}{12}\selectfont \textcolor{black}{$^{16}$O}}}
		\put(-50,55){{\fontsize{12}{12}\selectfont \textcolor{black}{$^{20}$Ne}}}
		\end{picture}		
		\caption{The values of $\chi^2$/NDF for the best estimates of the cluster parameters for VMC, NLEFT, PGCM, and 3pF are presented. The parameters can be found in the first and third rows of Table \ref{tab1}.. The $\chi^2$/NDF values serve as a criterion for selecting the sampled nucleons. The values corresponding to the nucleon configurations for neon are indicated within the shaded area.} 
		\label{Fige3}
	\end{figure}
	The fluctuations of $E$ can be found by the variance of total energy by truncating at the first nontrivial order in the perturbation, $\delta \varepsilon(\mathbf{r})$.
	The variance is obtained using a two-point correlation function $\la\varepsilon(\mathbf{r})\varepsilon(\mathbf{r}')\ra$ of Eq.\ref{A8}: 
	\begin{equation}\label{var}
	\begin{split}
	\text{var}(E/\la E\ra)&=\frac{\la E^2\ra_{ev}-\la E\ra_{ev}^2}{\la E\ra_{ev}^2}\\&=\frac{\int d\Omega d\Omega'\int d^2\mathbf{r}\;d^2\mathbf{r}'\;\la\delta\epsilon(\mathbf{r},\Omega,\Omega')\delta\epsilon(\mathbf{r}',\Omega,\Omega')\ra}{\Big(\int d\Omega\;d\Omega'\;\la\epsilon(\mathbf{r},\Omega,\Omega')\ra\Big)^2},
	\end{split}
	\end{equation}
	where  $\la\delta\epsilon(\mathbf{r})\delta\epsilon(\mathbf{r}')\ra=\la\epsilon(\mathbf{r})\epsilon(\mathbf{r}')\ra-\la\epsilon(\mathbf{r})\ra\la\epsilon(\mathbf{r}')\ra$ denotes the connected two-point function of the density field \cite{Blaizot:2014nia}. To verify the validity of Eq.\ref{var} from a field theory standpoint, I derive the correlation function of the energy field $\varepsilon(\mathbf{r})$ by computing the second derivative of the generating functional, or partition function, given by $k_2(\mathbf{r},\mathbf{r}')=\Big[\delta^2 W[j]/\delta j(\mathbf{r})\delta j(\mathbf{r}')\Big]\Big |_{j=0}$, with respect to the source field $j(\mathbf{r})$. This process facilitates the explicit calculation of the 2-point correlation function. In Appendix \ref{app3}, I demonstrate that this approach yields the variance as follows:
	\begin{equation}
	\begin{split}
	\text{var}(E)=\int d^2\mathbf{r}_1 d^2\mathbf{r}_2\; \Big[&k_2^{(1\times 2)}+k_2^{(\text{same}\times\text{same})}\\&+k_2^{(\text{diff}\times\text{diff})}+2k_2^{(\text{same}\times\text{diff})}\Big].
	\end{split}
	\end{equation}
	This formulation shows that $\alpha$ clustering affects the initial fireball not through the mean energy density, but entirely through its fluctuations. The one-body (mean-field) contribution cancels from second cumulant, so all sensitivity to nuclear substructure enters through two-body correlations. These naturally split into contributions from nucleons belonging to the same $\alpha$ cluster, which encode the intrinsic size and quantum structure of the $\alpha$, and contributions from nucleons belonging to different $\alpha$ clusters, which encode the special arrangement of the clusters inside the nucleus. Their mixed terms describe the interference between internal cluster structure and inter-cluster geometry.
	
	The geometry of the overlap zone is formulated by the initial anisotropy $\mathcal{E}_2$. To calculate the mean-square anisotropy, I consider the leading expression as follows:
	\begin{equation}\label{eps2}
	\varepsilon_2\{2\}^2\equiv\la \mathcal{E}_2\mathcal{E}_2^*\ra_{ev}=\frac{\int_{\mathbf{r}}\int_{\mathbf{r}'}|\mathbf{r}|^2|\mathbf{r}'|^2e^{2i(\phi-\phi')}\la\delta\epsilon(\mathbf{r})\delta\epsilon(\mathbf{r}')\ra_{ev}}{\left(\int_{\mathbf{r}} |\mathbf{r}|^2\la\epsilon(\mathbf{r})\ra_{ev}\right)^2}.
	\end{equation} 
	As an example, the forms of Eqs.\ref{var} and \ref{eps2} can be written for the spherical ion collisions as 
	\footnote{If we rewrite $\text{var}(E/\la E\ra)$ and $\la \mathcal{E}_2\mathcal{E}_2^*\ra_{ev}$ in the spherical case using the ratio $a\equiv w/R_s$ and expand to the first order of $a$, we have:
		\begin{align*}
		\text{var}(E/\la E\ra)&\approx 
		A^{-2}((1/6a^2)-(11/12))+A^{-1}(2/3),\\
		\varepsilon_2\{2\}^2&\approx 
		A^{-2}((1/3a^2)-(155/54))+A^{-1}(128/54).
		\end{align*}
		These calculations diverge in the limit of point particles ($w\to0$). Actually, this issue appears in $\mathcal{I}_{1,i}$ when we calculate the part of diagonal elements of  $\la\epsilon(\mathbf{r},\Omega,\Omega')\epsilon(\mathbf{r}',\Omega,\Omega')\ra$. This is because I have assumed a Gaussian distribution for distributed nucleons within a cluster (or a nucleus). However, it works correctly for non-vanishing $w$.}:
	\begin{align}
	\text{var}(E/\la E\ra)&=\frac{R_s^8+2(1+A_s)(2R_s^6w^2+3R_s^4w^4)}{3A_s^2w^2(2R_s^2+3w^2)(R_s^4+8R_s^2w^2+12w^4)},\label{varS}
	\end{align}
	\begin{align} 
	\begin{split}
	\varepsilon_2\{2\}^2&=\frac{16}{27A_s^2}\Big(\frac{4(A_s-1)(R_s^3+3R_sw^2)^4}{(R_s^4+8R_s^2w^2+12w^4)^3}\\&\hspace*{3cm}+\frac{9R_s^4}{16R_s^2w^2+24w^4}\Big).
	\end{split}\label{eps2S}
	\end{align}
	 
	As can be seen, these calculations allow us to study the effects of different nuclear parameters on the observables. After finding the forms of observables, it is time to check the validity of the calculations concerning different nucleon configurations of oxygen and neon obtained by VMC, NLEFT, and PGCM approaches (labeled original configurations (OC)). I also repeat my calculations for completeness with the configurations of $^{16}$O and $^{20}$Ne obtained by starting from a 3pF density distribution parameterized via $R_0=2.608$ fm, $a_0=0.513$ fm, $w=-0.051$ for $^{16}$O, and $R_0=2.791$ fm, $a_0=0.698$ fm, $w=-0.168$ for $^{20}$Ne \cite{ANGELI201369}. I also label these configurations by OC. In the next section, I obtain the initial correlators regarding O+O, Ne+Ne, Pb+O, and Pb+Ne collisions concerning various OC. To reproduce these configurations utilizing a cluster model, I use the methods introduced in Refs.\cite{Rybczynski:2017nrx} such that first I find the positions ($\vec{L}_i$) and the root-mean-square radii ($r_L$) of the clusters by finding the best estimations (labeled cluster configurations (CC)) for the radial distributions of the sampled nucleons in the theories as mentioned earlier. To do so, I choose the best estimations concerning the Pearson's $\chi^2$ test. By replacing the cluster parameters in Eqs.\ref{A4}, \ref{var}, and \ref{eps2} and taking numerical integration on Euler angles, the observables are resulted. To check the validity of my calculations, I compare the analytical results with the results obtained by the TRENTo simulations and the Monte Carlo code introduced in Ref.\cite{Giacalone:2023hwk}.       
	\begin{figure*}[t!]
		\begin{tabular}{c}
			\includegraphics[scale=.45]{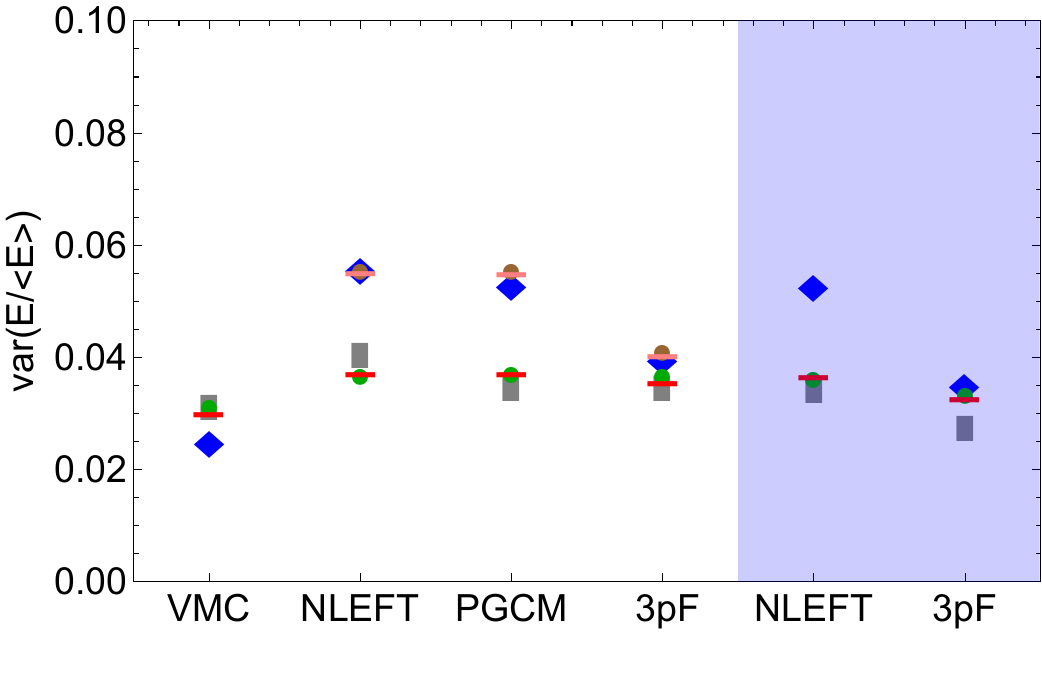}
			\includegraphics[scale=.45]{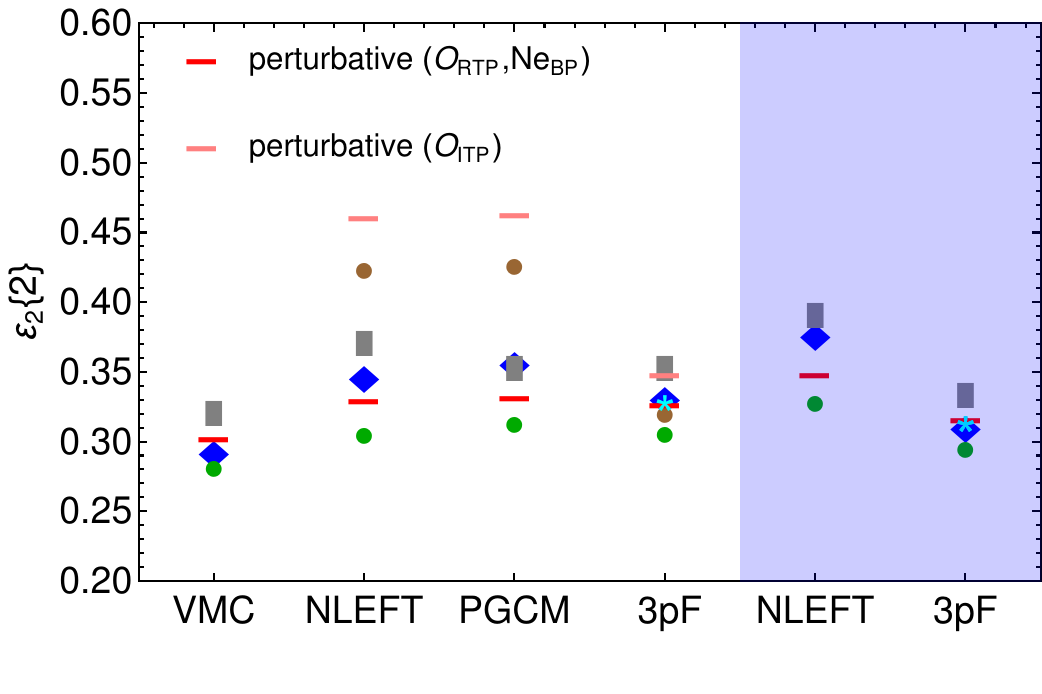}\\
			\includegraphics[scale=.45]{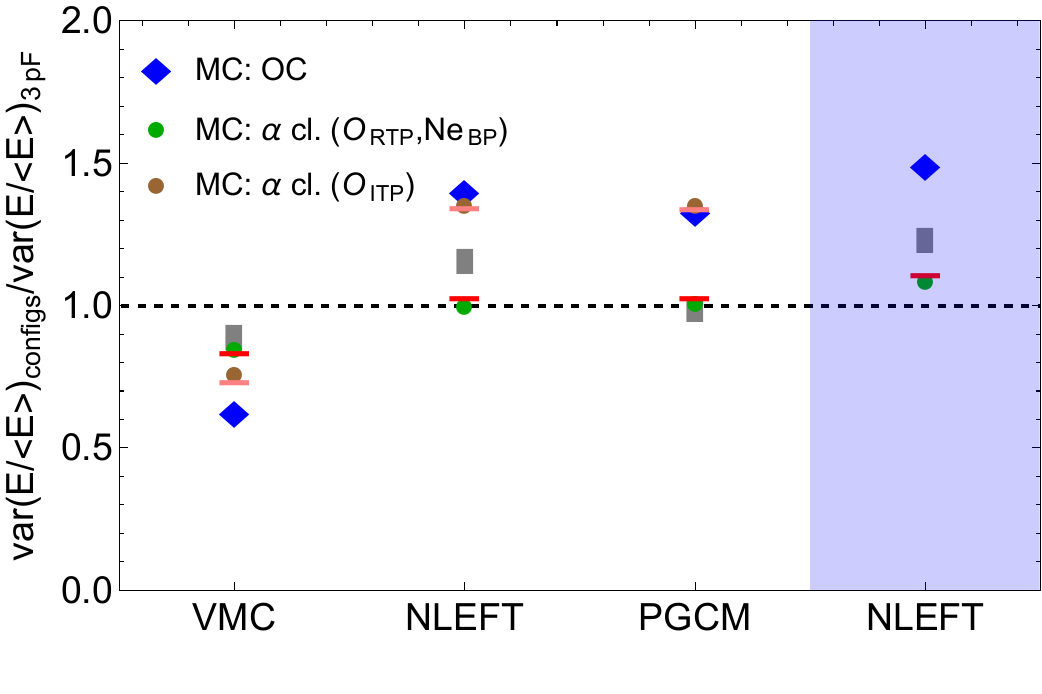}
			\includegraphics[scale=.45]{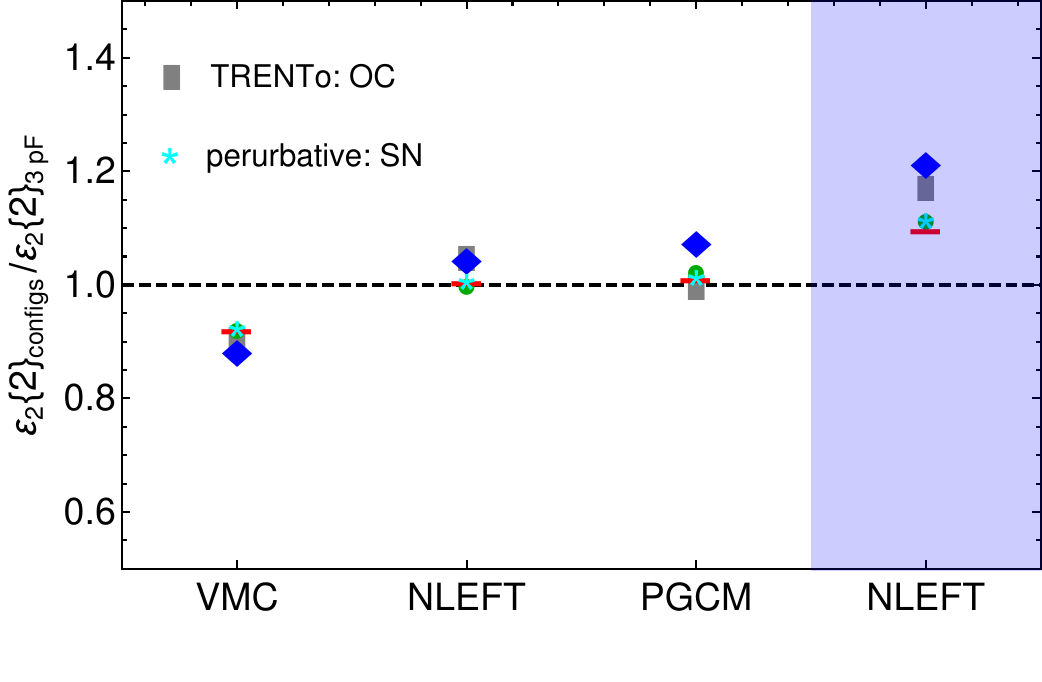}\\
			\includegraphics[scale=.45]{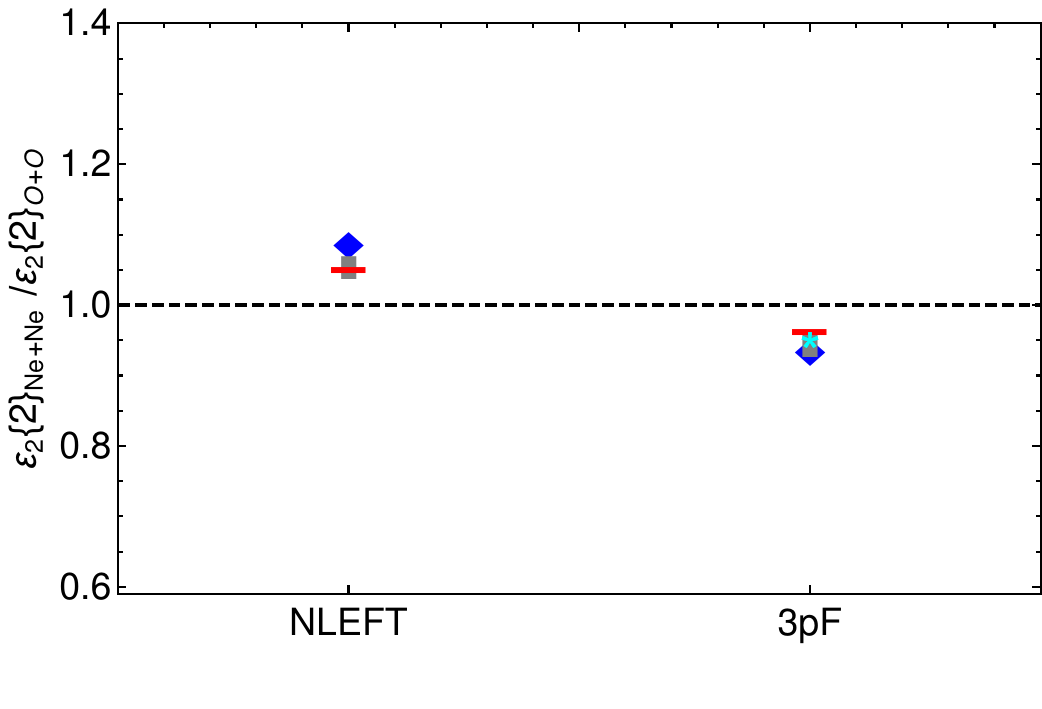}
			\includegraphics[scale=.45]{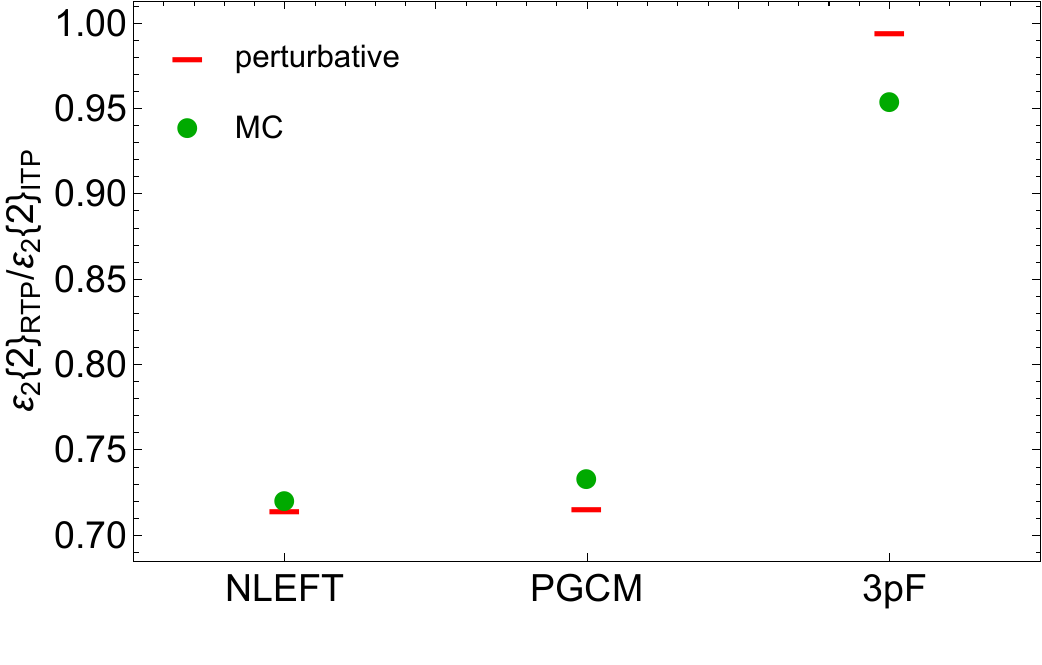}
		\end{tabular}
		\begin{picture}(0,0)
		\put(-260,130){{\fontsize{6}{6}\selectfont \textcolor{black}{$\times0.32$}}}
		\put(-425,115){{\fontsize{11}{11}\selectfont \textcolor{black}{$(a)$}}}
		\put(-195,115){{\fontsize{11}{11}\selectfont \textcolor{black}{$(b)$}}}
		\put(-425,-40){{\fontsize{11}{11}\selectfont \textcolor{black}{$(c)$}}}
		\put(-195,-40){{\fontsize{11}{11}\selectfont \textcolor{black}{$(d)$}}}
		\put(-425,-195){{\fontsize{11}{11}\selectfont \textcolor{black}{$(e)$}}}
		\put(-195,-195){{\fontsize{11}{11}\selectfont \textcolor{black}{$(f)$}}}
		\put(-290,210){{\fontsize{11}{11}\selectfont \textcolor{black}{$^{20}$Ne+$^{20}$Ne}}}
		\put(-395,210){{\fontsize{11}{11}\selectfont \textcolor{black}{$^{16}$O+$^{16}$O}}}
		\put(-60,-180){{\fontsize{11}{11}\selectfont \textcolor{black}{$^{16}$O+$^{16}$O}}}
		\end{picture}		
		\caption{The results of initial correlators, var($E/\la E\ra$) and $\varepsilon_2\{2\}$, for collisions involving $^{16}$O and $^{20}$Ne are presented in panel (a) and (b). 		
			The results of Ne+Ne collisions are highlighted by the shaded area in panels (a) through (d). Additionally, the ratios of $\mathcal{O}_{\text{configs}}/\mathcal{O}_{\text{3pF}}$ for different models of the oxygen and neon structures are shown in panels (c) and (d). I identify three types of results: perturbative calculations (represented by red and pink horizontal lines), Monte Carlo results (depicted by green and brown circles), and the results from TRENTo simulations at $\sqrt{s_{NN}}=200$ GeV (illustrated by gray rectangles) for O+O and Ne+Ne collisions. Note that TRENTo model results in panel (a) have been multiplied by 0.32 for better comparison. The results from Eqs.\ref{varS} and \ref{eps2S} for spherical shapes of oxygen and neon are indicated by cyan starts. The relative variation of $\varepsilon_2\{2\}$ between O+O and Ne+Ne  collisions is presented in panel (e).
			The difference of regular (Fig.\ref{Fige5}(a)) and irregular (Fig.\ref{Fige5}(b)) triangular pyramid shapes for oxygen in $\varepsilon_2\{2\}$ are illustrated in panel (f). The values of cluster parameters can be found in Table \ref{tab1}. Note that statistical error bars are smaller than the symbols used in the figures.} 
		\label{Fige4}
	\end{figure*}
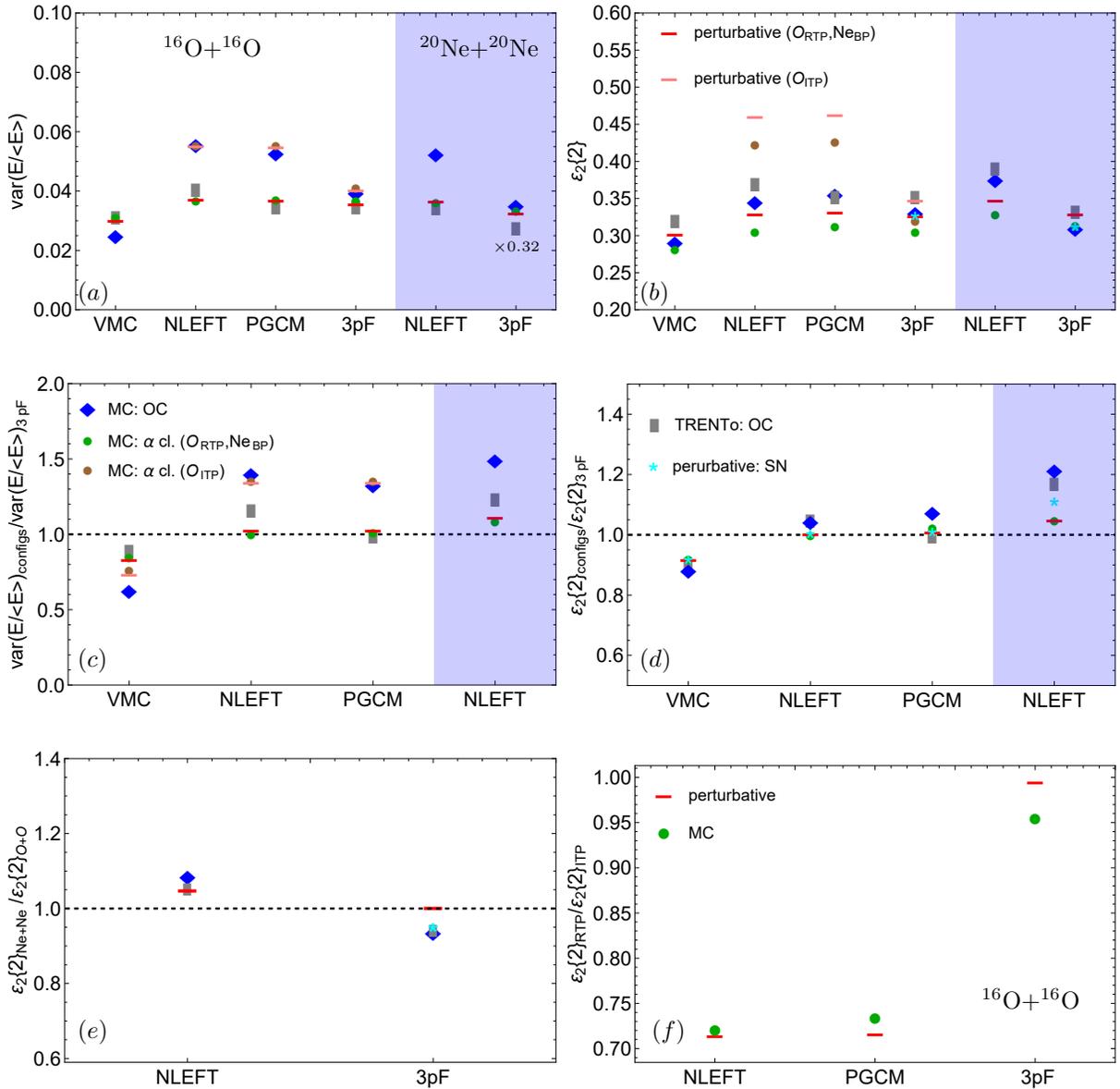
	
	\section{Results and discussion}\label{res}
	In this section, I study the properties of OC computed by VMC, NLEFT, and PGCM for $^{16}$O and NLEFT for $^{20}$Ne within a cluster model proposed in Ref.\cite{Rybczynski:2017nrx}. In this way, I investigate two different types of collisions, symmetric ($^{16}$O+$^{16}$O and $^{20}$Ne+$^{20}$Ne) and asymmetric ($^{208}$Pb+$^{16}$O and $^{208}$Pb+$^{20}$Ne). This investigation can be studied for other light nuclei to enhance our understanding of capsuled features in the nucleon configurations of light nuclei related to different Hamiltonians or models. The approach of this paper contains two steps:
	\begin{enumerate}
		\item Finding positions and sizes of clusters regarding one-body density distributions\footnote{I should also mention that to obtain the positions of nucleons, I considered the configurations that incorporate information from the two-body density distribution using the method introduced in Ref.\cite{Liu:2025uks}.}.
		\item Analytical calculations and comparison with Monte Carlo results (see Secs.\ref{sy} and \ref{asy}).
	\end{enumerate}    
\textit{Cluster features}: In the first step, I find the nucleon positions inside of each cluster by employing Eq.\ref{q1} and nucleon possibility distribution,
\begin{align*}
P(r)\equiv 4\pi r^2 \rho_N(\vec{r})=4\pi r^2\sum_i^{N_{\alpha}} \rho_{\alpha_i}(\vec{r}),
\end{align*}  
in the framework proposed by the authors in Ref.\cite{Rybczynski:2017nrx}\footnote{As I mentioned, I modified this framework to capture the two-body density distribution as well (see Appendix \ref{app1}).}. Then I compare the nuclear one-body density of CC \footnote{Since I am working in zero impact parameter collisions, I found the positions of cluster centers such that the center-of-mass of sampled nucleons are in the origin of the 3D coordinate system.} with OC to find the best description concerning Pearson's $\chi^2$ test. In this way, I search all $\chi^2$ minima, and then I accept CC that has the closest $\chi^2$/NDF (or reduced Chi-squared) values to 1. Fig.\ref{Fige2} shows the best descriptions (red points) for the configurations sampled according to PGCM, NLEFT, VMC, and 3pF for oxygen in panels (a)-(d) and NLEFT and 3pF for Ne in panels (e) and (f). Regarding the explanations above, it can be understood that the depicted red (point) curves in Fig. \ref{Fige2} were not obtained through fitting. Instead, the error bars in each bin represent the differences between the results derived from various $\chi^2$ minima, while the points indicate the $\chi^2$ minima that are closest to 1.
The values of true $\chi^2$/NDF for different structures are depicted in Fig.\ref{Fige3}. As can be seen, the results indicate the best descriptions ($\chi^2$/NDF$\approx1$) of the nucleon configurations computed by \textit{ab initio} models and 3pF distributions. I should mention that here I generated the sampled nucleons based on tetrahedron or Regular Triangular Pyramid (RTP) shapes for oxygen and a BP structure for neon (see the first and third rows of Table \ref{tab1}), as the shapes are illustrated in panel Fig.\ref{Fige1}(a). From now on, I discuss the results based on the distance of two cluster centers instead of the cluster positions. Table \ref{tab1} shows the obtained estimations of the cluster parameters ($r_L$,$\ell_c$) for oxygen and ($r_L$,$\ell_c$,$\ell_h$) for neon (see Fig.\ref{Fige1}(a)). I should note that the displayed parameters here are related to the minimum $\chi^2$ value that is closest to 1.
Having these values, I will study the effects of $\alpha$ clusters for $^{16}$O \footnote{These parameters satisfy the experimental root-mean-square radius, which is approximately 2.7 fm.} and $^{20}$Ne on the initial correlators in the next parts.

\begin{figure}[t!]
\begin{tabular}{c}
		\hspace*{-.9cm}\includegraphics[scale=.66]{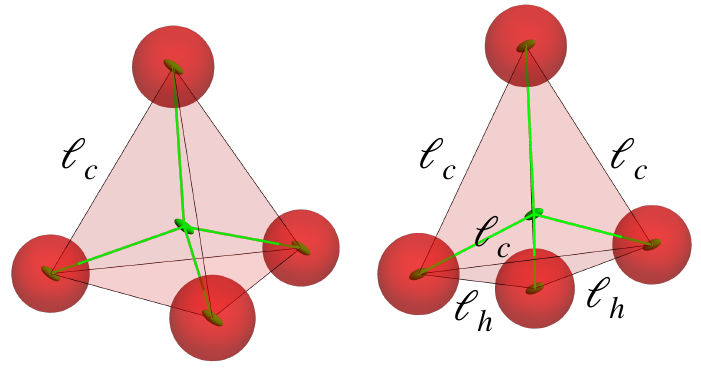}
\end{tabular}
\begin{picture}(0,0)
	\put(-70,-65){{\fontsize{13}{13}\selectfont \textcolor{black}{ITP}}}
	\put(-175,-65){{\fontsize{13}{13}\selectfont \textcolor{black}{RTP}}}
	\put(-230,45){{\fontsize{13}{13}\selectfont \textcolor{black}{(a)}}}
	\put(-120,45){{\fontsize{13}{13}\selectfont \textcolor{black}{(b)}}}
\end{picture}		
\caption{The schematics of regular (a) and irregular (b) triangular pyramids for oxygen are presented. The values of cluster parameters for different configurations are shown in the second and third rows of Table \ref{tab1}.} 
\label{Fige5}
\end{figure}

\subsection{Symmetric Collisions}\label{sy}
Relativistic O+O and Ne+Ne collisions provide a high potential to study the existence of quark-gluon plasma in the collisions of light nuclei as well as intrinsic structure properties of oxygen and neon where $\alpha$-particle description is relevant. Recently, these collisions have potentially been studied such that they have been proposed for the upcoming LHC runs \cite{Giacalone:2024ixe,Giacalone:2024luz,Mehrabpour:2025rzt}. Since we use the nucleon configurations obtained by \textit{ab initio} models to study these types of collisions, extracting encoded information in these samples is required. This can be reached by describing the configurations within a cluster model \footnote{Notice that I denoted the configurations obtained by the cluster model for oxygen and neon with $\alpha$ cl. in all figures.}. In this section, I try to figure out the features of OC by investigating the effects of cluster parameters, $r_L$ and $\ell_i$, on the initial correlators in comparison with the results of stated theories. Fig.\ref{Fige4}(a) and (b) present the results of var($E/\la E\ra$) and $\varepsilon_2\{2\}$ for $^{16}$O+$^{16}$O and $^{20}$Ne+$^{20}$Ne collisions (shaded area). Fig.\ref{Fige4} shows three types of results obtained by MC simulation (circles and diamonds) introduced in Ref.\cite{Giacalone:2023hwk}, perturbative calculations (horizontal lines) presented in Sec.\ref{corr}, and TRENTo model (gray rectangles) \footnote{Here I did not consider the dependency of the MC and perturbative calculations to the center-of-mass energy $\sqrt{s_{NN}}$ of the collisions.}. The blue diamonds indicate the results of OC using the MC simulation. I should mention that the correlators are calculated regarding RTP (or tetrahedron) and BP shapes for oxygen and neon, respectively. The MC results of CC and the perturbative calculations for these shapes are illustrated by green circles and red horizontal lines. As can be seen, MC and perturbative results have similar values of var($E/\la E\ra$). This means that all MC information can be captured by perturbative calculations for a cluster model. However, the results of OC and CC obtained by NLEFT and PGCM for oxygen as well as NLEFT for neon are not consistent, but the same tendency is followed. 
	
I show that if we find var($E$) concerning an irregular triangular pyramid (ITP) shape for oxygen \cite {Bijker:2020} (brown circles), we can capture the OC results obtained by MC simulations. The ratios in Fig.\ref{Fige4}(c) show clearly this status. The perturbative calculations of ITP are displayed by pink horizontal lines. Nevertheless, the study of initial anisotropy $\mathcal{E}_2$ for OC indicates an agreement with the tetrahedron shapes of oxygen, as illustrated in panel (b). To see the difference between RTP and ITP structures, the results of $\varepsilon_2\{2\}_{\text{RTP}}/\varepsilon_2\{2\}_{ITP}$ are presented in Fig.\ref{Fige4}(f) for CC. This figure demonstrates a comparison between RTP and ITP shapes. The same behavior can be seen from perturbative calculations and MC results approximately. The difference in estimations of NLEFT and PGCM is more than $25\%$. It is worth mentioning that $\varepsilon_2\{2\}$ obtained starting from the 3pF charge density marks a near RTP shape. In the case of neon, since there are several possibilities to find the same results with var($E$) for NLEFT, I do not consider them in this study. The results of $\varepsilon_2\{2\}$ show that a BP shape for neon can approximately describe NLEFT results of $^{20}$Ne.  
	
	\begin{table}[t!]
		\caption{The table presents the $\alpha$-cluster parameters for oxygen ($r_L$,$\ell_c$) and for neon ($r_L$,$\ell_c$,$\ell_h$). In addition to the RTP shapes for oxygen, I also investigate the ITP shapes (see Fig.\ref{Fige5}) for NLEFT, PGCM, and 3pF. I found that the smallest length $\ell_h$ should be approximately $0.52 \ell_c$ for both NLEFT and PGCM, while it should be about $0.9 \ell_c$ for 3pF. The parameters ($r_L$,$\ell_c$) for the ITP shape are shown in the second row of the table.}
		{\footnotesize \begin{supertabular}{| l | c c c c |}
				\hline
				Models & 3pF & NLEFT & PGCM & VMC \\ \hline
				$^{16}$O (RTP) & (1.83,3.20) & (1.84,3.17)& (1.88,3.06) & (1.52,3.26) \\ \hline
				$^{16}$O (ITP)& (2.00,3.15) & (1.61,3.88) & (1.57,3.86) & -\\	\hline
				$^{20}$Ne (BP)& (2.00,3.00,3.00) & (2.20,3.00,3.50) & - & -\\ \hline
		\end{supertabular}}
		\label{tab1}
	\end{table}
The initial correlators obtained for the different configurations are normalized to the results derived by the sampled nucleons concerning 3pF charge density distribution. I study this ratio to determine whether the results of original and cluster configurations indicate independent or dependent sampling nucleons \cite{Zhang:2024vkh,Lu:2025cni}. The results of ratios are illustrated in Figs.\ref{Fige4}(c) and (d). The correlators mark the correlated nucleons for the configurations derived by \textit{ab-initio} models. The smaller values for RTP structures are because I find approximately the same cluster parameters for NLEFT, PGCM, and 3pF ($r_L/\ell_c\approx0.6$) such that the:
	\begin{align*}
	\text{MC}:\frac{var(E/\la E\ra)_{\text{NLEFT}}}{var(E/\la E\ra)_{\text{3pF}}}&\approx 1.000, \frac{\varepsilon_2\{2\}_{\text{NLEFT}}}{\varepsilon_2\{2\}_{\text{3pF}}}\approx 0.998,\\
	\text{pert}:\frac{var(E/\la E\ra)_{\text{NLEFT}}}{var(E/\la E\ra)_{\text{3pF}}}&\approx 1.043, \frac{\varepsilon_2\{2\}_{\text{NLEFT}}}{\varepsilon_2\{2\}_{\text{3pF}}}\approx 1.009,
	\end{align*}
	\begin{align*}
	\text{MC}:\frac{var(E/\la E\ra)_{\text{PGCM}}}{var(E/\la E\ra)_{\text{3pF}}}&\approx 1.011, \frac{\varepsilon_2\{2\}_{\text{PGCM}}}{\varepsilon_2\{2\}_{\text{3pF}}}\approx 1.023,\\
	\text{pert}:\frac{var(E/\la E\ra)_{\text{PGCM}}}{var(E/\la E\ra)_{\text{3pF}}}&\approx 1.041, \frac{\varepsilon_2\{2\}_{\text{PGCM}}}{\varepsilon_2\{2\}_{\text{3pF}}}\approx 1.015, 
	\end{align*}
	where "pert" denotes the perturbative calculations. However, we can see the perturbative calculations for variance var($E/\la E\ra$) show $4\%$ deviations approximately. Of course, the results of $\varepsilon_2\{2\}$ show a consistency between MC and analytical calculations.
	
	For a complement study, I generate $10^6$ events for $^{16}$O+$^{16}$O and $^{20}$Ne+$^{20}$Ne collisions at 200 GeV in zero impact parameter using the TRENTo model. I use the nucleon configurations derived by the \textit{ab initio} theories as well as 3pF distributions. TRENTo results shown by the gray rectangles in Fig.\ref{Fige4} indicate the tendency of TRENTo outputs to be followed by the calculations performed on the cluster configurations (MC and perturbative calculations). Notice that I scaled the TREENTo results by a factor of 0.32 in panel (a) for better comparison. This difference in variance comes from where I did not consider the $\sqrt{s_{NN}}$ in MC and perturbative calculations. The ratios in panels (c) and (d) also indicate a qualitative agreement between TRENTo and CC results. However, in the selections of the nucleon configurations, in addition to the one-body density, the nucleon-nucleon correlations are required to capture all information of light nuclear structures \cite{Rybczynski:2019adt}.
	
	Computing the Ne+Ne/O+O ratio, shown in Fig.\ref{Fige4}(e), highlights instead the strong impact of the neon shape concerning the NLEFT model, which enhances $\mathcal{E}_2$ by more than 5\%. Also, the results shows larger size diminishes $\varepsilon_2\{2\}$ if we compare the 3pF results \cite{Giacalone:2024ixe}. We can see there is a consistency in NLEFT and 3pF results. Moreover, for the nucleon configurations obtained by starting 3pF distributions, I use  Eq.\ref{Qsph} to find a spherical description. To have the best prediction, I tried to find $R_s$ concerning the similar values for var($E/\la E\ra$) indicated by the blue diamonds. It leads to $R_s = 2.252$ fm and $2.63$ fm for oxygen and neon respectively. This allows us to control both one- and two-body correlations simultaneously. Then, I calculated the values of $\varepsilon_2\{2\}_{\text{3pF}}$ as denoted by cyan stars in Fig.\ref{Fige4}(b), (d) and (e). One can find  $\frac{\varepsilon_2\{2\}_{\text{configs.}}}{\varepsilon_2\{2\}_{\text{3pF}}}\approx \frac{\varepsilon_2\{2\}_{\text{configs.}}}{\varepsilon_2\{2\}_{\text{sph}}}$ for the oxygen and neon structures. After all this, the ratio $\frac{\varepsilon_2\{2\}_{\text{Ne+Ne}}}{\varepsilon_2\{2\}_{\text{O+O}}}$ shows a good agreement with the results of OC obtained by both TRENTo and MC simulations.

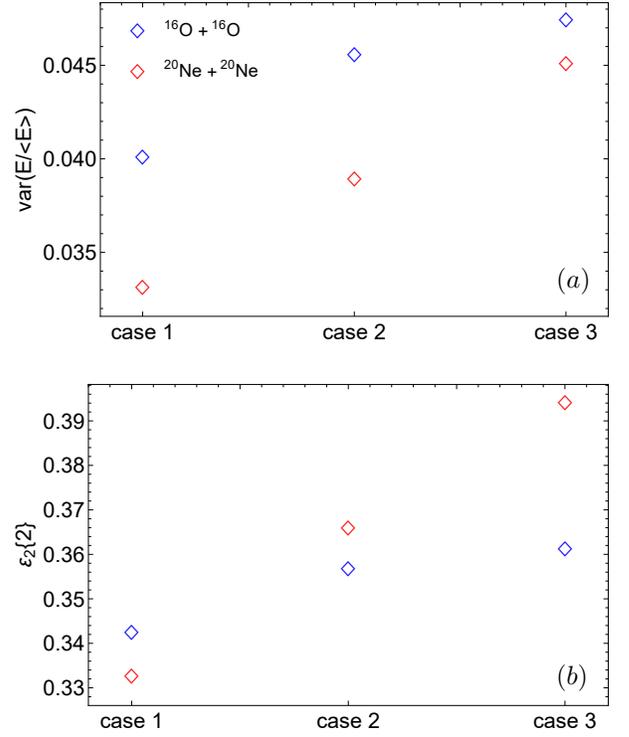
\begin{figure}[t!]
		\begin{tabular}{c}
			\includegraphics[scale=.45]{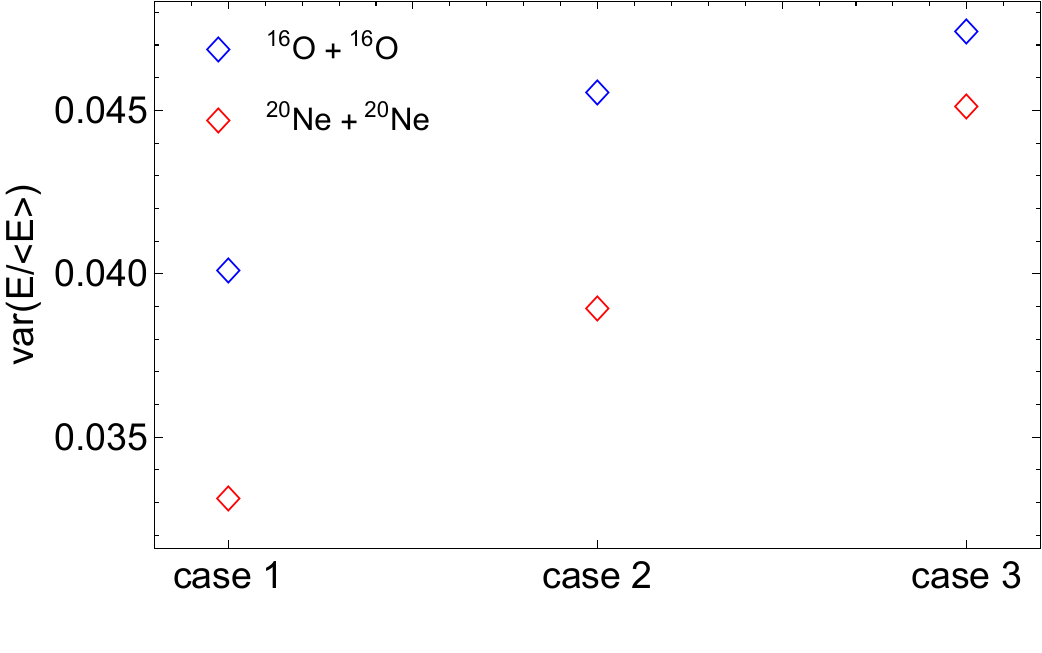}\\
			\includegraphics[scale=.45]{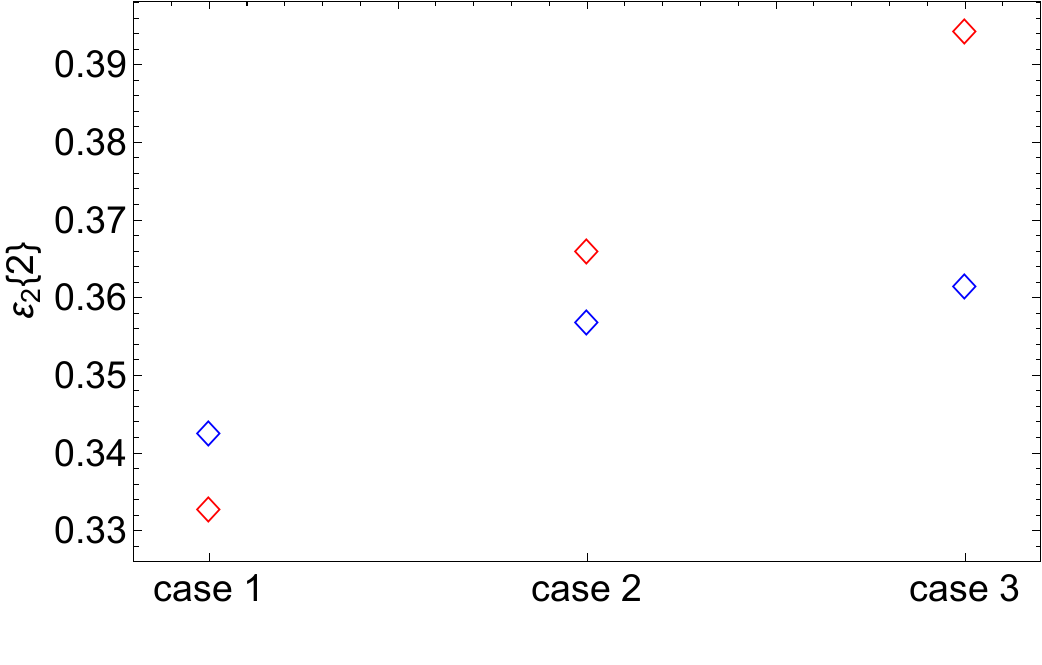}
		\end{tabular}
		\begin{picture}(0,0)
		\put(-25,40){{\fontsize{11}{11}\selectfont \textcolor{black}{$(a)$}}}
		\put(-25,-110){{\fontsize{11}{11}\selectfont \textcolor{black}{$(b)$}}}
		\end{picture}		
		\caption{The initial correlators results, var($E/\la E\ra$) (a) and $\varepsilon_2\{2\}$ (b), for the collisions of $^{16}$O+$^{16}$O and $^{20}$Ne+$^{20}$Ne are presented by blue and red diamonds, respectively. These results are obtained from perturbative calculations related to the RTP structure (case 2) and $\alpha$+ET structures (case 1 and 3) for oxygen, as well as the RTB structure (case 1) and BP structures (case 2 and 3) for neon. The cluster parameters are detailed in Table \ref{tab2}.} 
		\label{Fige6}
	\end{figure}

	\begin{table}[t!]
		\caption{The table presents the values of $\alpha$-cluster parameters for various shape states of oxygen and neon. The structures are categorized based on the relationships between parameters: for the $\alpha$ + Equilateral-Triangle (ET) configuration, I assume that $\ell_c<\ell_h$, while for the RTP structure, I set $\ell_c=\ell_h$. Additionally, I explore the case where $\ell_c>\ell_h$ for a smashed tetrahedron originating from one of its vertices. Furthermore, I examine the BP and Regular Trigonal Bipyramid (RTB) shapes for 5-$\alpha$ clusters, considering both scenarios of $\ell_c<\ell_h$ and $\ell_c=\ell_h$.}
		\begin{supertabular}{| c | c   c   c |}
			\hline
			\quad structures \quad& case 1 & case 2 & case 3\\ \hline
			\quad4-$\alpha$ clusters\quad\quad & \quad$\ell_h=\ell_c/2$ \quad& \quad$\ell_h=\ell_c$ \quad&\quad $\ell_h=3\ell_c/2$\quad\quad  \\ \hline
			\quad5-$\alpha$ clusters\quad\quad& $\ell_h=\ell_c$ & $\ell_h=3l_c/2$ & $\ell_h=2l_c$  \\ \hline
		\end{supertabular}
		\label{tab2}
	\end{table}
Recently, the ratio of $r_L/\ell_c$ has been studied for $^{16}$O \cite{YuanyuanWang:2024sgp,Shafi:2025feq}. To do so, they enforced that the root-mean-square radius $\sqrt{<r^2>}=\sqrt{ ((3/8)\ell_c^2+r_L^2)}$ of $^{16}$O should be the same for the different densities. Yet various shapes have not been investigated. Here, I study the different cases for $^{16}$O (RTP and $\alpha$ + Equilateral-Triangle (ET) \cite{Furutachi:2007vz}) and $^{20}$Ne (RTB \cite{Bijker:2020gbl,Adachi:2020gjp} and $\alpha$+RTP or BP \cite{Ebran:2017rcb}) structures.
The cluster parameters ($r_L,\ell_c$) for different shapes can be found in Table \ref{tab2}. 
Notice that here in the choice of parameters, I don't enforce to get the same $\la r^2\ra$ for the different densities. The results of var($E/\la E\ra$) are demonstrated in Fig.\ref{Fige6}(a) as well as $\varepsilon_2\{2\}$ is depicted in panel (b). The initial correlators are enhanced by increasing the values of $\ell_h$, and this trend is followed for both O+O and Ne+Ne collisions. The results indicate that we have $\mathcal{O}_{\alpha+\text{ET}} > \mathcal{O}_{\text{RTP}} > \mathcal{O}_{\text{smashed RTP}}$ as well as $\mathcal{O}_{\text{BP}} > \mathcal{O}_{\text{RTB}}$, as can be seen in Fig.\ref{Fige6}, such that case 1/case 2 $\approx 88\%$ for var(E/⟨E⟩) in $^{16}$O, case 2/case 3 $\approx  96\%$ for var(E/⟨E⟩) in $^{16}$O, case 1/case 2 $\approx 96\%$ for $\varepsilon_2\{2\}$ in $^{16}$O and case 2/case 3 $\approx 99\%$ for $\varepsilon_2\{2\}$ in $^{16}$O. Therefore, the difference between the ratios shows the enhancement slope of correlators in O+O collisions is decreasing for large $\ell_h$. 
When the length scale  $\ell_h$  significantly exceeds $\ell_c$, both the $\alpha$+ET configuration in four clusters and the BP configuration in five clusters effectively approximate two distinct regions of point matter. Under these conditions, the size of the cluster can be considered negligible, as expected. However, based on the constraints imposed by the experimental radius of the root mean square, it is clear that  $\ell_h$  cannot exceed  $\ell_c$. Therefore, in practical scenarios, the size of the cluster must be taken into account.
	
	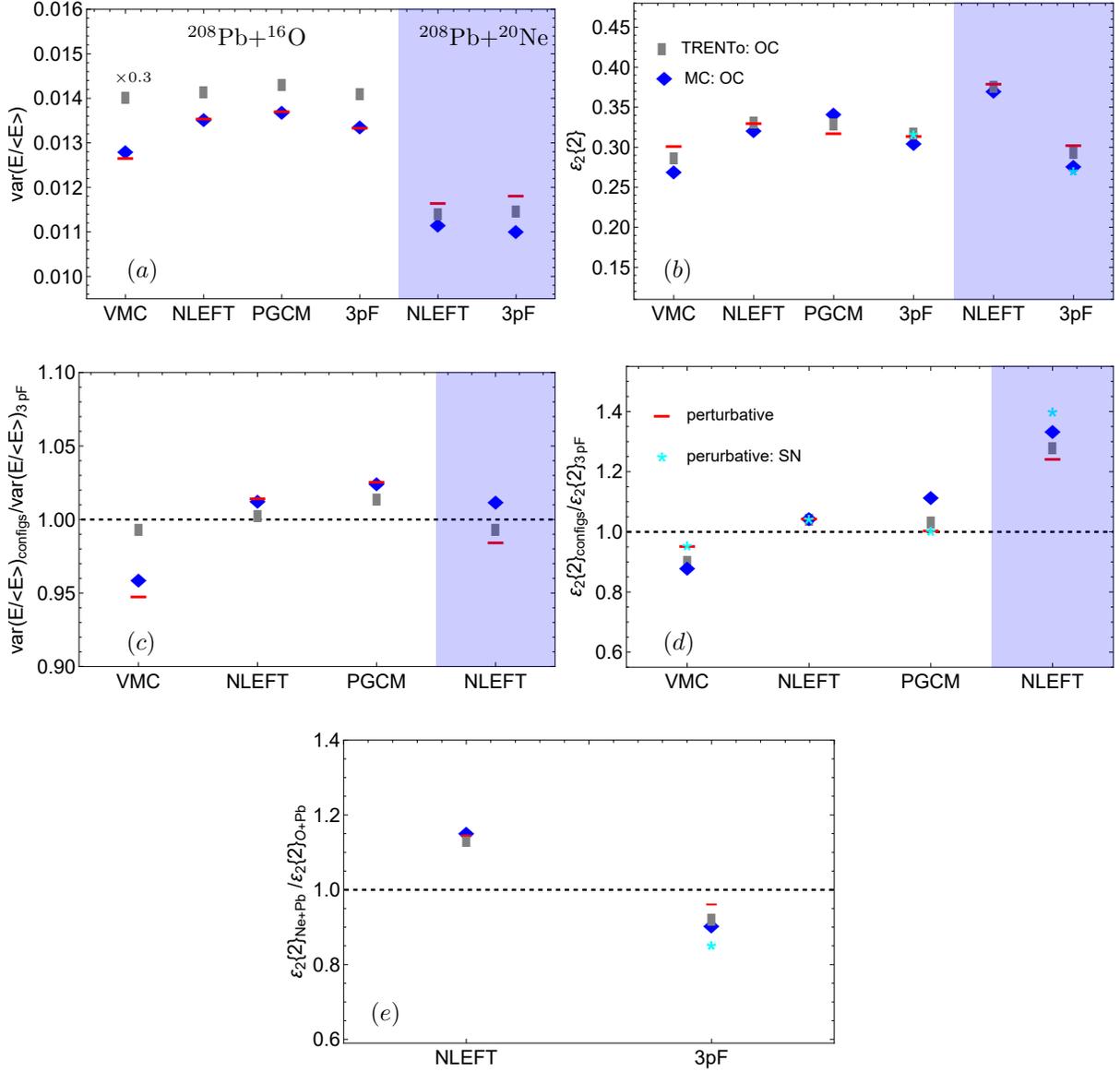
\begin{figure*}[t!]
		\begin{tabular}{c}
			\includegraphics[scale=.45]{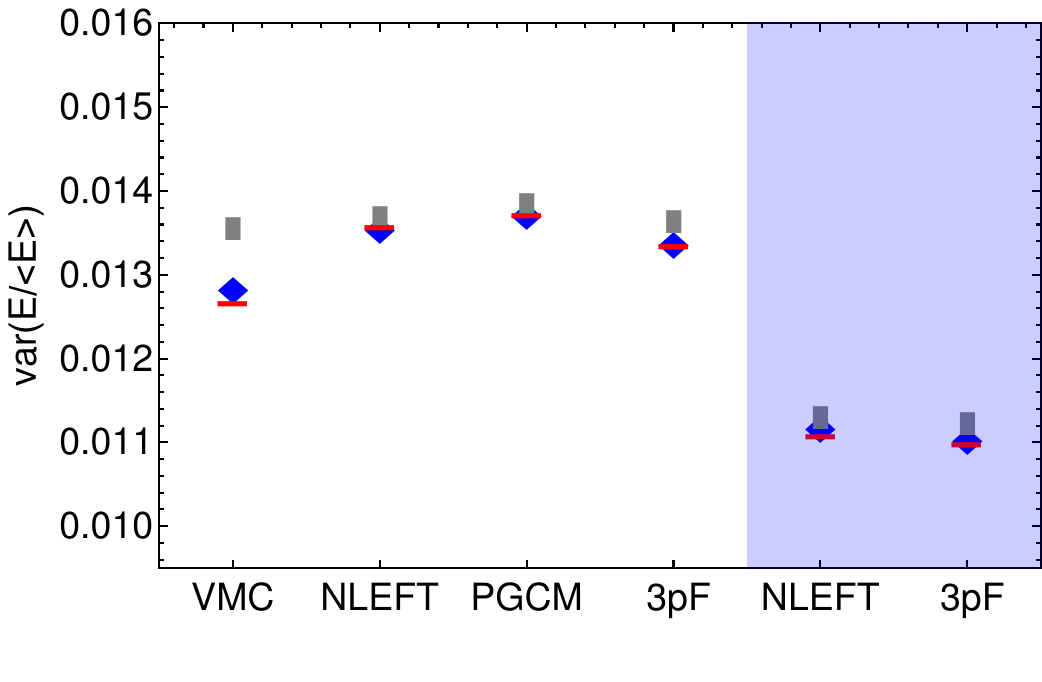}
			\includegraphics[scale=.45]{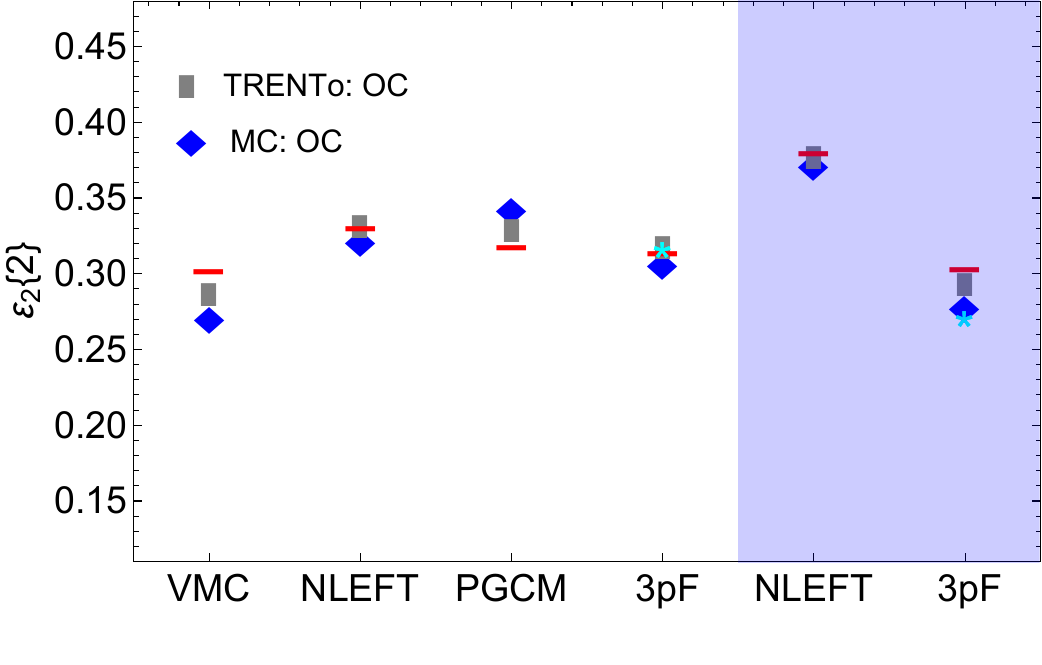}\\
			\includegraphics[scale=.45]{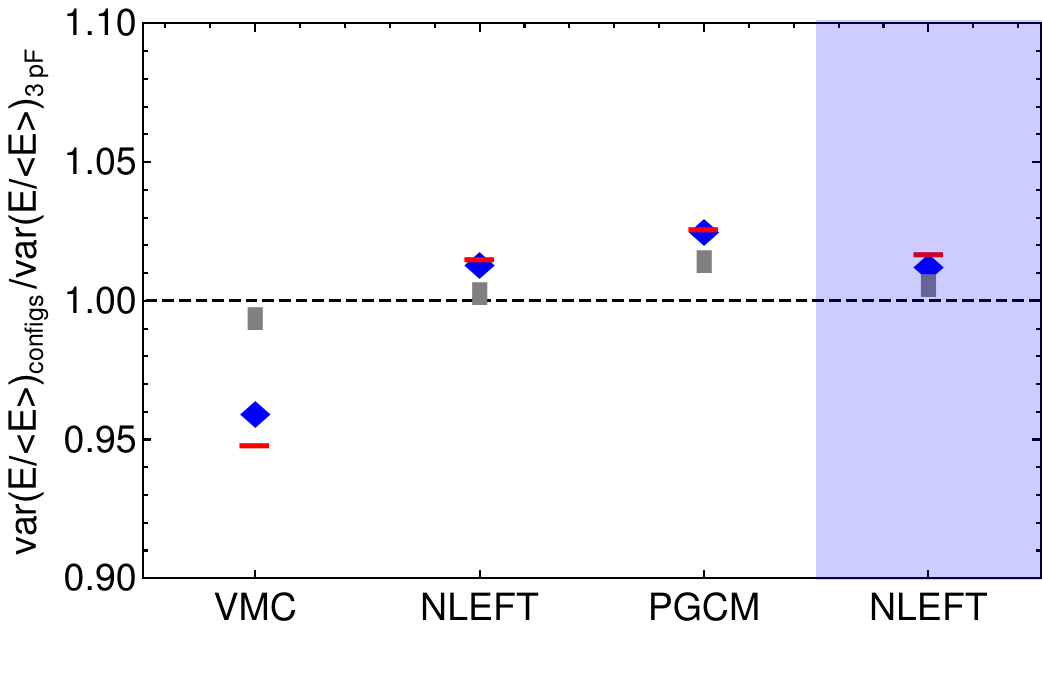}
			\includegraphics[scale=.45]{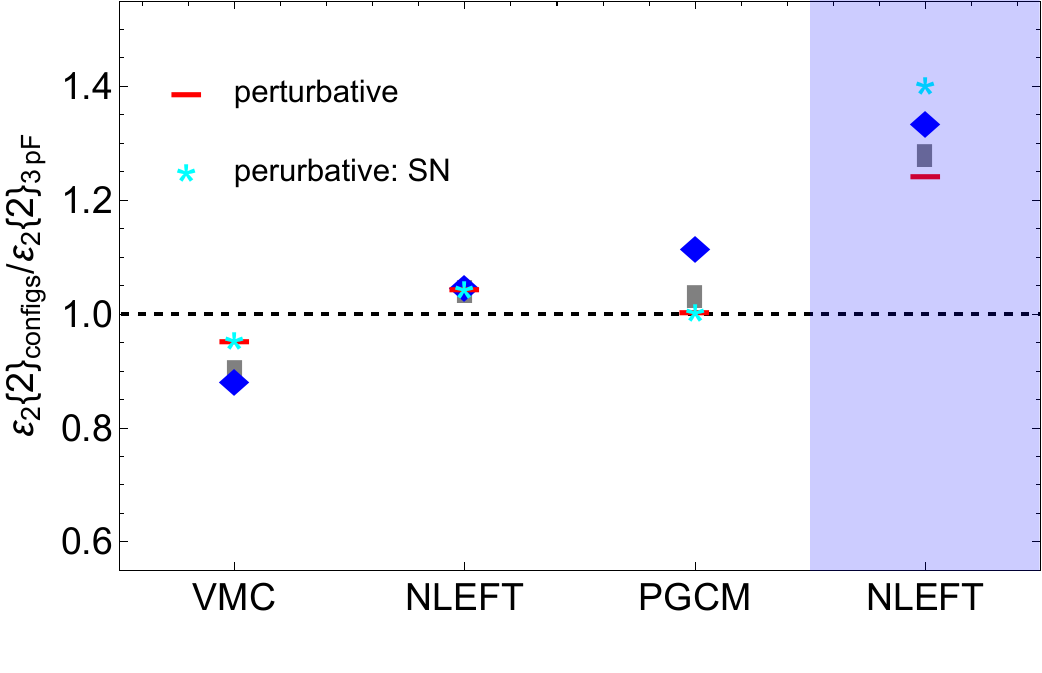}\\
			\includegraphics[scale=.45]{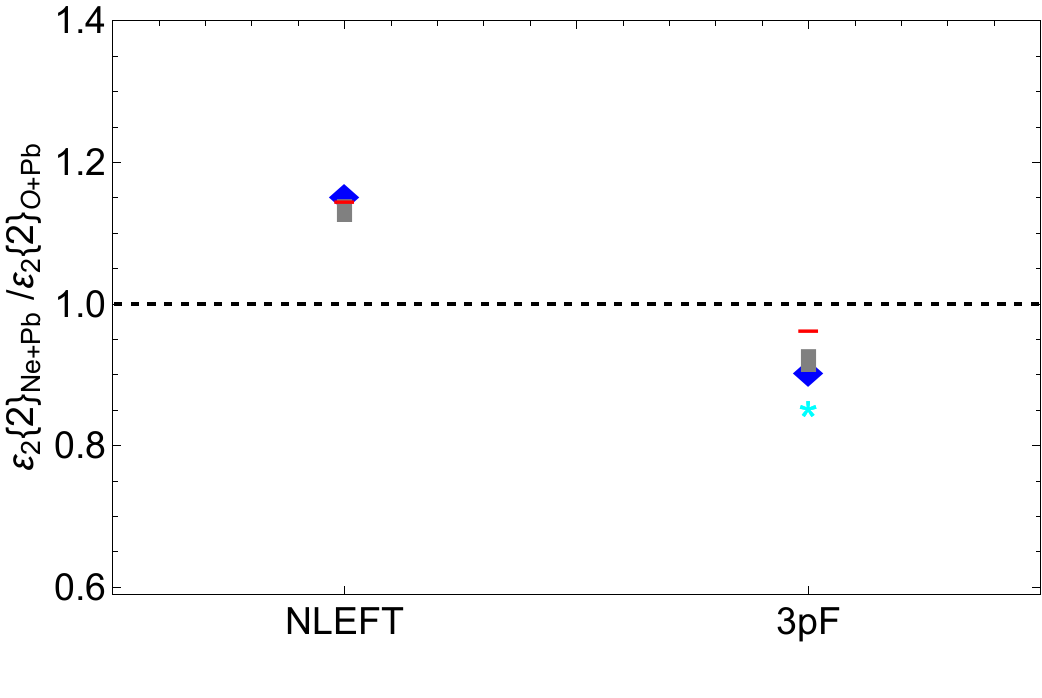}
		\end{tabular}
		\begin{picture}(0,0)
		\put(-415,195){{\fontsize{6}{6}\selectfont \textcolor{black}{$\times0.3$}}}
		\put(-410,115){{\fontsize{11}{11}\selectfont \textcolor{black}{$(a)$}}}
		\put(-190,115){{\fontsize{11}{11}\selectfont \textcolor{black}{$(b)$}}}
		\put(-410,-38){{\fontsize{11}{11}\selectfont \textcolor{black}{$(c)$}}}
		\put(-190,-38){{\fontsize{11}{11}\selectfont \textcolor{black}{$(d)$}}}
		\put(-310,-190){{\fontsize{11}{11}\selectfont \textcolor{black}{$(e)$}}}
		\put(-295,210){{\fontsize{11}{11}\selectfont \textcolor{black}{$^{208}$Pb+$^{20}$Ne}}}
		\put(-385,210){{\fontsize{11}{11}\selectfont \textcolor{black}{$^{208}$Pb+$^{16}$O}}}
		\end{picture}		
		\caption{Similar to Fig.\ref{Fige4}, this figure presents results for the collisions of $^{208}$Pb+$^{20}$Ne and  $^{208}$Pb+$^{16}$O. Additionally, these asymmetric collisions were simulated using the TRENTo model at an energy of $68.5$ GeV with zero impact parameter. To facilitate a clearer comparison, the results from the TRENTo model in panel (a) have been scaled by a factor of 0.3.} 
		\label{Fige7}
	\end{figure*}
	
	\subsection{Asymmetric Collisions}\label{asy}
	One of the main plans in a fixed target experiment at SMOG system of the LHCb detector \cite{LHCb:2021ysy} is a direct study of light nuclear structures, such as $^{16}$O and $^{20}$Ne, involves colliding them with heavy spherical ions, such as $^{208}$Pb \cite{Broniowski:2013dia,Rybczynski:2017nrx}. This way provides data with higher statistics to scrutinize the light nuclear structures \cite{LHCb:2021ysy}. In this part, I study the collisions of $^{208}$Pb+$^{16}$O and $^{208}$Pb+$^{20}$Ne to check the validity of perturbative calculations for the different structures of oxygen and neon. Here, I consider various nucleon configurations of oxygen obtained by NLEFT, PGCM, VMC, and 3pF. I also consider the configurations generated by NLEFT and 3pF for $^{20}$Ne. In this way, the average of the total system's energy density in Eq.\ref{q20} should be written as follows:
	\begin{equation}
	\begin{split}
	\la E\ra_{ev}&=\frac{3A\times B}{2\pi(r_L^2+R_s^2+6w^2)}
	\\&\hspace*{1.5cm}\times\sum_{i=1}^{N_{\alpha}}\int d\Omega\; e^{f_i+\frac{(h_{x,i}^2+h_{y,i}^2)r_L^2(R_s^2+6w^2)}{6(r_L^2+R_s^2+6w^2)}},
	\end{split}	
	\end{equation}     
	where the functions $f_i$, $h_{x,i}$ and $h_{y,i}$ are read from Eqs.\ref{fi}-\ref{hy}. We note that it is considered all nucleons of light ions have participated in the collisions. Instead, a range for the number of participants from Pb should be supposed. It is because, in a realistic model, a part of nucleons from Pb only participated in this type of collision. Therefore, I find the range $\Delta A=A_{\text{max}}-A_{\text{min}}$ by generating Pb+O and Pb+Ne collisions at $\sqrt{s_{NN}}=68.5$ GeV in zero impact parameters using a 2D TRENTo model. Of course, the asymmetric collisions studied in the fixed target experiments require 3D simulations, but I want to compare 2D data with the results of MC simulations and mentioned perturbative calculations to have an apples-to-apples comparison. The results of MC and TRENTo simulations concerning OC are depicted by the blue diamonds and gray rectangles in Fig.\ref{Fige7}, respectively. Since the event-by-event fluctuations of the number of participants ($N_{\text{part}}$) are described by a Gaussian distribution, to find the correct perturbative results averaging over the weighted correlators are required \footnote{Notice that I used $e^{-(A-\bar{A})^2/2\sigma^2}$ instead the normalized to avoid finding the correlators very small.}:
	\begin{equation}\label{modeqs}
	\mathcal{O} = \frac{1}{\Delta A}\sum_{A=A_{\text{min}}}^{A_{\text{max}}} e^{-(A-\bar{A})^2/2\sigma^2} \mathcal{O}(A), 
	\end{equation}
	where $\bar{A}=\la N_{\text{part}}\ra$ and $\sigma=\la(N_{\text{part}}-\la N_{\text{part}}\ra)^2\ra^{1/2}$ are found from the TRENTo results. This allows us to have an apples-to-apples comparison with simulation results.  
	Concerning $\bar{A}_{\text{Pb+O}}/\bar{A}_{\text{Pb+Ne}}\approx84.4\%$ and $\sigma_{\text{Pb+O}}/\sigma_{\text{Pb+Ne}}\approx94.5\%$ \footnote{The values of ($\bar{A},\sigma$) obtained by TRENTo are approximately (65,8.6) and (77,9.1) for oxygen and neon, respectively. Concerning these values, I considered the range $\Delta A\approx3\sigma$ because this range gives better predictions of correlators.}, I calculate the weighted initial correlators by replacing var($E/\la E\ra$) and $\varepsilon_2\{2\}$ in Eq.\ref{modeqs}. The red horizontal lines in Fig.\ref{Fige7} illustrate the results of perturbative calculations. Notice that I only considered the RTP or tetrahedron shapes for oxygen configurations, as well as the BP for neon structures, in my calculations.
	To describe the one-nucleon distribution concerning charge density of $^{208}$Pb, I use a same method, as I did for spherical oxygen and neon, such that I considered a Wood-Saxon (WS) distribution, $N_{\text{WS}}(r) \propto \big(1+\exp[(r-R_0)/a_0]\big)^{-1}$, for $^{208}$Pb and then I found var($E/\la E\ra$) using data generated by MC simulation on Pb+Pb collisions. Utilizing Eg.\ref{Qsph}, I found $R_s=3.84$ fm for $^{208}$Pb. By replacing $R_s$ in Eq.\ref{eps2S} and taking a square root, I obtained $\frac{\varepsilon_2\{2\}_{\text{perturbative}}}{\varepsilon_2\{2\}_{\text{MC-WS}}}\approx 90\%$. Although the difference is large, the obtained parameters work well in the collisions which are involving Pb+O and Pb+Ne. 
	
	As illustrated in Fig.\ref{Fige7}(a), there is a strong consistency between the perturbative calculations (represented by the red horizontal lines) and the Monte Carlo (MC) simulations (depicted as blue diamonds) for for both of oxygen and neon structures. To facilitate comparison, the results from the TRENTo model for variance were scaled by a factor of 0.3. Notably, the trend indicated by the gray rectangles aligns well with the perturbative calculations.
	In Fig.\ref{Fige7}(c), I present the ratio $\text{var}(E/\la E\ra)_{\text{configs}}/\text{var}(E/\la E\ra)_{\text{3pF}}$. Here, a significant gap is observed between the results from TRENTo and those obtained from the starting Variational Monte Carlo (VMC) method. However, the trend results are completely consistent with those illustrated in Fig.\ref{Fige4}.
	Fig.\ref{Fige7}(b) further demonstrates that there is consistency between the perturbative calculations and simulations regarding the  $\varepsilon_2\{2\}$ results. The ratios shown in panel (d) exhibit excellent agreement between analytical calculations and TRENTo outputs.
	To assess the impact of neon structures, I computed the ratio $\varepsilon_2\{2\}_{\text{Pb+Ne}}/\varepsilon_2\{2\}_{\text{Pb+O}}$, which serves as a discriminator, as shown in panel (e). This figure reveals that the effects of neon configurations derived from NLEFT and 3pF are captured by both perturbative calculations and MC simulations. 
	The cyan stars in Fig.\ref{Fige7} signify that both colliding nuclei are spherical. As depicted in panels (b) and (d), there is a correspondence between the red lines and cyan stars for oxygen, indicating unity, while the results differ for neon. This discrepancy is further illustrated in panel (e), where I investigate the shape impact of neon in comparison to oxygen.

	\section{Conclusion}\label{conclusion}
	To investigate light nuclear structures, including $\ alpha$ clusters, I employed a Gaussian distribution to arrange the nucleons within the clusters of light nuclei. Once the nucleon distribution was established, I analytically studied the two-point correlators to demonstrate the impact of cluster parameters using a rotor model. This analysis enabled us to explore both symmetric collisions (O+O and Ne+Ne) and asymmetric collisions (Pb+O and Pb+Ne) such that I showed that this approach effectively extracts the characteristics of various \textit{ab-initio} models within an $\alpha$-cluster framework, by examining the structures of $^{16}$O and $^{20}$Ne. We note that these calculations provide a general framework using a cluster model for studying the structures of light ions. This comprehensive analysis highlights the intricate relationship between cluster parameters and collision dynamics in these nuclei. To determine the cluster parameters, I minimized the $\chi^2$ statistic to identify the distributions that best aligned with the configurations predicted by \textit{ab-initio} models such as NLEFT, PGCM, and VMC, as well as a three-parameter Fermi (3pF) charge density function. From the 3pF results, I derived an analytical spherical description for oxygen and neon. The findings indicated that perturbative calculations effectively capture the features of the 3pF model in the ratios of $\varepsilon_2\{2\}_{\text{configs}}/\varepsilon_2\{2\}_{3pF}$. Additionally, by comparing with spherical $^{208}$Pb using a Wood-Saxon density to generate nucleon positions, I calculated $R_s$ to derive observables from X+Pb collisions. 
	
	To validate my perturbative calculations, I compared the results with Monte Carlo (MC) simulations and TRENTo outputs. The findings demonstrate a good agreement between the trends observed by initial correlators in the perturbative calculations and those predicted by the TRENTo model. This indicates that perturbative calculations can quantitatively capture the structural features of different Hamiltonians/structures concerning light nuclear collisions. However, to fully understand the behavior of correlators in asymmetric collisions, it is crucial to consider the weighted initial correlators. The variance var($E$) in symmetric collisions shows that the VMC is more compatible with a tetrahedral shape when compared to the MC results for oxygen configurations, while other methods suggest an irregular triangular pyramid structure. These findings also confirm the presence of an $\alpha$-cluster structure resembling a distinct bowling-pin shape for neon configurations computed using the NLEFT approach. In this work, I examined different structural states for $^{16}$O (RTP and $\alpha$ + ET) and $^{20}$Ne (RTB and $\alpha$+RTP). The calculations revealed a relationship among observables such that $\mathcal{O}_{\alpha+\text{ET}}>\mathcal{O}_{\text{RTP}}>\mathcal{O}_{\text{smashed RTP}}$ and $\mathcal{O}_{\text{BP}}>\mathcal{O}_{\text{RTB}}$. Furthermore, calculations indicate that when analyzing structures such as $\alpha$ + a lighter ion (for example, $\alpha$ + $^{12}$C for oxygen and $\alpha$ + $^{16}$O for neon), the length of the side $\ell_h$ cannot substantially exceed $\ell_c$. In these cases, the size of the cluster is not negligible, which challenges our initial assumptions.
	
	Since fixed-target collisions provide a clearer environment for studying the structures of light nuclei, I repeated my calculations for Pb+O and Pb+Ne collisions. To accurately reconstruct realistic asymmetric collisions, I demonstrated that it is necessary to weight the number of participating nucleons from heavy spherical nuclei. The calculations showed good consistency with the Monte Carlo (MC) and TRENTo results of the original configurations concerning the weighted correlators. Furthermore, the results indicate that perturbative calculations can capture more information in asymmetric collisions.

	\section*{Acknowledgments}
	I thank Giuliano Giacalone, Matthew Luzum, Chun Shen, Li Yan, Xu-Guang Huang, Abhisek Saha, and Shuzhe Shi for fruitful discussions. I also acknowledge the valuable conversations with the participants of the 4th international workshop on QCD collectivity at the smallest scales, where this work was initiated, as well as those at Quark Matter 2025. This research is partially supported by the National Natural Science Foundation of China under Grant No. 12247107.

\appendix
\section{Two-body density distribution}\label{app1}
To examine whether the presented analytical approach can reconstruct nucleon-nucleon correlations, I first check the two-body density of the system:
\begin{equation}
\begin{split}
\\&\rho^{(2)}(r_1, \theta_1, \phi_1, r_2, \theta_2, \phi_2) \\&=\frac{1}{8\pi^2}\int_{\Omega}\rho_N(R_{xzx}(\Omega)(r_1, \theta_1, \phi_1))\rho_N(R_{xzx}(\Omega)(r_2, \theta_2, \phi_2)),
\end{split} 
\end{equation}
which is obtained from the angular average of the two-point function of the intrinsic density. If the intrinsic density in Eq.\ref{q4} is spherical, we have:
\begin{equation}\label{aq0} 
\rho^{(2)}(r_1, \theta_1, \phi_1, r_2, \theta_2, \phi_2) = \rho^{(1)}(r_1, \theta_1, \phi_1)\rho^{(1)}(r_2, \theta_2, \phi_2).
\end{equation}
On the other hand, correlations arise in the terms related to cluster parameters. I will explain the broken equality in a simplified case. It is not straightforward to calculate the two-body distribution this way when we have two clusters located at a distance $2\ell$ from each other along the z-axis. The one-nucleon distribution within the clusters is described as follows:
\begin{align}\label{aq1}
\rho_{\alpha}(x,y,z) = \Big(\frac{3}{2\pi r_L^2}\Big)^{3/2} \Big(f_- (x,y,z) +f_+ (x,y,z)\Big),
\end{align}
where $$f_{\pm}(x,y,z)=\exp\Big[-\frac{3(x^2+y^2+(z\pm l)^2}{2r_L^2}\Big].$$
Therefore, after rotating Eq.\ref{aq1} with respect to the Euler angles, we obtain:
\begin{equation}\label{aq2}
\begin{split}
\rho_{\alpha}(x,y,z,\Omega) &= \rho_{\alpha}(R_{zxz}(\Omega)(x,y,z))
\\&=\Big(\frac{3}{2\pi r_L^2}\Big)^{3/2}\Big(f_{-}(x,y,z,\Omega) + f_{+}(x,y,z,\Omega) \Big),
\end{split}
\end{equation}
where 
\begin{equation}\label{aq3}
\begin{split}
f_{\pm}(x,y,z,\Omega) &= e^{-\frac{3(l^2+x^2+y^2+z^2)}{2 r_L^2}} \\&\times e^{\mp
	\frac{3l \big( x \sin (a_1) \sin (a_3)+ y \sin (a_2) \cos (a_3)+ z \cos (a_2)\big)}{r_L^2}}. 
\end{split}
\end{equation}
With $\rho_{\alpha}(x,y,z,\Omega)$ from Eq.\ref{aq2}, one can find the two-body density distribution as follows:
\begin{equation}\label{aq4}
\begin{split}
\rho_{\alpha}^{(2)}&(x_1,y_1,z_1;x_2,y_2,z_2)\\&=\frac{1}{8\pi^2} \int_{\Omega} \rho_{\alpha}(x_1,y_1,z_1,\Omega) \rho_{\alpha}(x_2,y_2,z_2,\Omega)
\\&= \Big(\frac{3}{2\pi r_L^2}\Big)^{3}\frac{1}{8\pi^2} \\&\hspace*{1cm}\times\int_{\Omega}\Big(f_{-}(x_1,y_1,z_1,\Omega) f_{-}(x_2,y_2,z_2,\Omega) \\&\hspace*{1.5cm}+ f_{-}(x_1,y_1,z_1,\Omega) f_{+}(x_2,y_2,z_2,\Omega) \\&\hspace*{1.5cm}+ f_{+}(x_1,y_1,z_1,\Omega) f_{-}(x_2,y_2,z_2,\Omega)
\\&\hspace*{1.5cm}+ f_{+}(x_1,y_1,z_1,\Omega) f_{+}(x_2,y_2,z_2,\Omega)\Big).
\end{split}
\end{equation}
Concerning Eq.\ref{aq3}, the forms of the above combinations are obtained as:
\begin{align*}
f_{-}(\vec{r}_1,\Omega) f_{-}(\vec{r}_2,\Omega)&=e^{A_1+A_2(z_1+z_2)+A_3(y_1+y_2)+A_4(x_1+x_2)},\\
f_{-}(\vec{r}_1,\Omega) f_{+}(\vec{r}_2,\Omega)&=e^{A_1+A_2(z_1-z_2)+A_3(y_1-y_2)+A_4(x_1-x_2)},\\
f_{+}(\vec{r}_1,\Omega) f_{-}(\vec{r}_2,\Omega)&=e^{A_1-A_2(z_1-z_2)-A_3(y_1-y_2)-A_4(x_1-x_2)},\\
f_{+}(\vec{r}_1,\Omega) f_{+}(\vec{r}_2,\Omega)&=e^{A_1-A_2(z_1+z_2)-A_3(y_1+y_2)-A_4(x_1+x_2)}.
\end{align*}
where 
\begin{align*}
A_1 &= -\frac{3}{2r_L^2}(2l^2+r_1^2+r_2^2),\\
A_2 &= \frac{3l}{r_L^2} \cos (a_2),\\
A_3 &= \frac{3l}{r_L^2} \cos (a_3) \sin (a_2),\\
A_4 &= \frac{3l}{r_L^2} \sin (a_3) \sin (a_2).
\end{align*} 
To demonstrate the broken equality in Eq.\ref{aq0}, it suffices to find the terms of Eq.\ref{aq4} in spherical coordinates. The evaluation of these integrals requires us to compute 
\begin{equation}
J(b_1,b_2)=\frac{1}{2}\int_{0}^{\pi}e^{b_1  \cos (a_2)}I_0\Big(b_2 \sin (a_2)\Big)\sin (a_2)\;da_2,
\end{equation}
where $I_0(x)$ denotes the modified Bessel function.
Although a neat closed form for this integral in elementary functions is not available, we can express it in a useful special-function or series form, and also provide representations that are well-suited for numerical work or asymptotic analysis. I will present only the final results as follows:
\begin{equation}\label{aq11}
J(b_1,b_2)= \frac{1}{2}\sum_{k=0}^{\infty}\frac{b_2^{2k}}{2^{2k}(k!)^2}\sqrt{\pi}(\frac{2}{b_1})^{k+1/2}\Gamma(k+\frac{3}{2})I_{k+\frac{1}{2}}(b_1),
\end{equation}
where $\Gamma(k+\frac{3}{2})=\frac{(2k+1)!!}{2^{k+1}}\sqrt{\pi}$.Therefore, using Eq.\ref{aq11}, we have:
\begin{align} 
\int_{\Omega} &f_{-}(\vec{r}_1,\Omega) f_{-}(\vec{r}_2,\Omega) =e^{A_1} J\Big(\frac{3l}{r_L^2}\xi_{+},\frac{3l}{r_L^2}\sqrt{\frac{1}{2}\kappa_{+}}\;\Big),\quad
\\
\int_{\Omega} &f_{-}(\vec{r}_1,\Omega) f_{+}(\vec{r}_2,\Omega) =e^{A_1} J\Big(\frac{3l}{r_L^2}\xi_{-},\frac{3l}{r_L^2}\sqrt{\frac{1}{2}\kappa_{-}}\;\Big),
\\ 
\int_{\Omega} &f_{+}(\vec{r}_1,\Omega) f_{+}(\vec{r}_2,\Omega) =e^{A_1} J\Big(-\frac{3l}{r_L^2}\xi_{+},\frac{3l}{r_L^2}\sqrt{\frac{1}{2}\kappa_{+}}\;\Big),
\\ 
\int_{\Omega} &f_{+}(\vec{r}_1,\Omega) f_{-}(\vec{r}_2,\Omega) =e^{A_1} J\Big(-\frac{3l}{r_L^2}\xi_{-},\frac{3l}{r_L^2}\sqrt{\frac{1}{2}\kappa_{-}}\;\Big).
\end{align}
where 
\begin{align*}
\xi_{\pm}&= r_1\cos\theta_1\pm r_2\cos\theta_2,\\
\kappa_{\pm} &= r_1^2+r_2^2-r_1^2\cos 2\theta_1-r_2^2\cos 2\theta_2\\&\hspace*{2cm}\pm 4r_1 r_2 \sin\theta_1\sin\theta_2\cos(\phi_1-\phi_2).
\end{align*}
It can be clearly concluded that the equality in Eq.\ref{aq0} is broken in this approach, as $\rho_{\alpha}^{(2)}(r_1,\theta_1,\phi_1;r_2,\theta_2,\phi_2) \neq \rho_{\alpha}^{(1)}(r_1,\theta_1,\phi_1) \rho_{\alpha}^{(1)}(r_2,\theta_2,\phi_2)$, which implies the existence of correlations between nucleons. Furthermore, one can extend these calculations to a greater number of clusters and arrive at the same conclusion in this approach. 

\textit{$-$Distance-distance correlations:} To assess the single-particle characteristics of the atomic ground state, one can drive a one-nucleon distribution that characterizes the nuclear structure using the one-body density matrix \cite{Alvioli:2009ab}, $\rho(\mathbf{r}) = A \int\prod_{i=2}^{A}d\mathbf{r}_i|\Psi(\mathbf{r},\mathbf{r}_2,\cdots,\mathbf{r}_A)|^2$, where $\Psi(\mathbf{r},\mathbf{r}_2,\cdots,\mathbf{r}_A)$ denotes the normalized $A$-nucleon ground-state wave function. 
It incorporates realistic short-range correlations, including $N$-body interactions, applied to an independent particle wave function \cite{Alvioli:2005cz,Pandharipande:1979bv}. To generate a one-body density that accounts for nucleon-nucleon interactions, Ref.\cite{Alvioli:2009ab} proposes a method based on the Metropolis algorithm  to search for configurations that satisfy the constrains imposed by two-body correlations,
$C(\mathbf{r},\mathbf{r}') = A (A-1) \int\prod_{i=3}^{A}d\mathbf{r}_i|\Psi(\mathbf{r},\mathbf{r}',\mathbf{r}_3,\cdots,\mathbf{r}_A)|^2$. This approach effectively retains the effects of short-range correlations on the single-particle density. The approximate form of $C(\mathbf{r},\mathbf{r}')$ is defined as follows: 
\begin{equation}
C(\Delta r) = 1 - \frac{\rho(\Delta r)}{\rho'(\Delta r)},\quad  \Delta r = |\mathbf{r}|,
\end{equation}	
where \(\mathbf{r}\) is the relative displacement. Here, the functions $\rho(\Delta r)$ and $\rho'(\Delta r)$ correspond  to the correlated and uncorrelated two-body densities, respectively \cite{Pieper:1992gr}. The function $C(\Delta r)$ intuitively captures the differences in distance between nucleons induced by nuclear forces. 

\begin{figure}[t!]
	\begin{tabular}{c}
		\includegraphics[scale=.48]{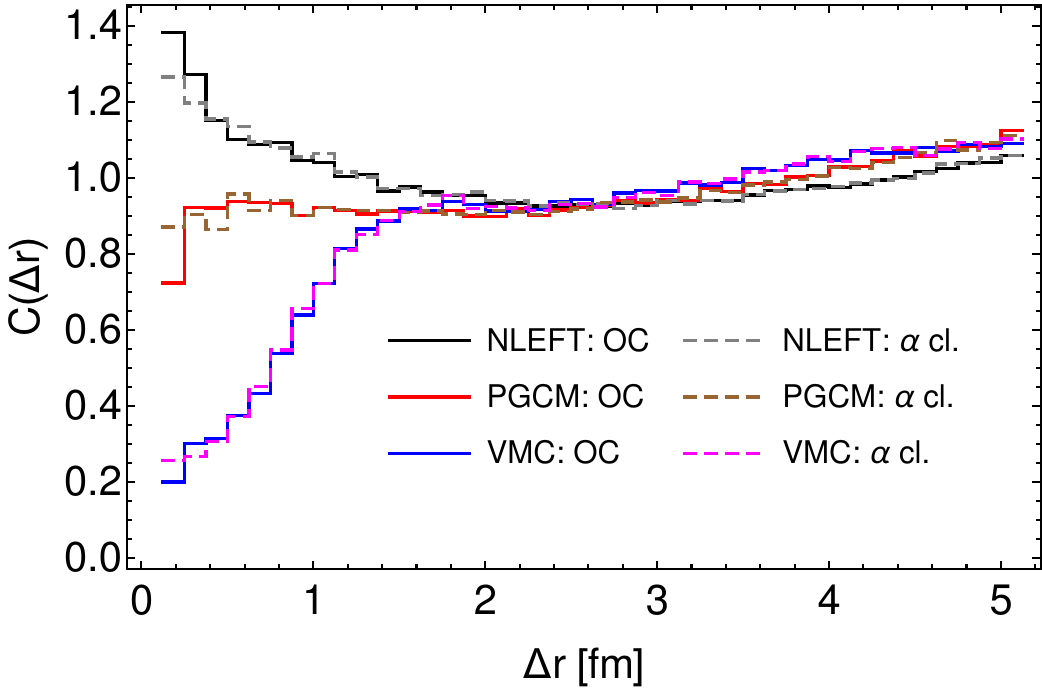}
	\end{tabular}
	\begin{picture}(0,0)
	\end{picture}		
	\caption{The two-body density distributions illustrated here are based on the one-body distributions presented in Fig.\ref{Fige2} (original (OC) and cluster ($\alpha$ cl.) configurations). I only present $C(\Delta r)$ for $^{16}$O; however, Similar results can be found for $^{20}$Ne as well.} 
	\label{Figetwo}
\end{figure}

To accurately represent distance-distance distributions of \textit{ab initio} models while determining the cluster parameters, I employ the \textit{acceptance-rejection method} framework \cite{Christensen}. This involves considering both one- and two-body densities computed from the OC to identify cluster parameters (refer to Ref.\cite{Liu:2025uks}). This methodology imposes constraints derived from two-body distribution of OC, thereby ensuring that the nucleon configurations are consistent with the OC.  As depicted in Fig.\ref{Figetwo}, the resulting $C(\Delta r)$ distributions demonstrate quantitative consistency with those computed using the OC.

\section{Two-point correlations}\label{app2}
To illustrate the form of the two-particle correlators and their dependence on the cluster parameters, I calculate the variance of the total energy density of systems in the overlap region\footnote{This would be straightforward if one wanted to show $\varepsilon_2\{2\}$ from Eq.\ref{eps2} in this manner. However, due to its complex form, I do not present it here.},
\begin{equation}
\text{var}(E/\la E\ra)= \frac{\la E^2\ra_{ev}}{\la E\ra_{ev}^2} -1= \frac{\int_{\Omega}\int_{\Omega'}\la E(\Omega,\Omega')^2\ra}{\int_{\Omega}\int_{\Omega'}\la E(\Omega,\Omega')\ra^2} -1
\end{equation}
where $\Omega$ and $\Omega'$ are considered for the rotations of nucleus A and nucleus B, respectively.
If we consider both colliding nuclei are the same, we obtain the average of total energy density by keeping the Euler angles as following:
\begin{equation}\label{qb2}
\la E(\Omega,\Omega')\ra = \frac{3A^2}{4\pi r_L^2(1+3\xi)}\sum_{i,j=1}^{N_{\alpha}}e^{\frac{3\eta(\Delta_{0,1}^{ij}(\Omega,\Omega')+\xi\Delta_{0,2}^{ij}(\Omega,\Omega'))}{24(1+3\xi)}}, 
\end{equation}
where $\xi=w^2/r_L^2$ and $\eta=\ell_c^2/r_L^2$.  
The functions of $\Delta^{ij}$ are defined as follows:
\begin{align*}
\Delta_{0,1}^{ij}&(\Omega,\Omega')=\frac{96 r_L^2}{\ell_c^2}\Big(f_i(\Omega)+f_j(\Omega')\Big)\\&+\frac{8r_L^2}{\ell_c^2}\Big(( h_{x,i}(\Omega)+h_{x,j}(\Omega'))^2+( h_{y,i}(\Omega)+h_{y,j}(\Omega'))^2\Big),\\
\Delta_{0,2}^{ij}&(\Omega,\Omega')= \frac{288 r_L^2}{\ell_c^2}\Big(f_i(\Omega)+f_j(\Omega')\Big)\\&+\frac{48r_L^2}{\ell_c^2}\Big( h_{x,i}(\Omega)^2+h_{x,j}(\Omega')^2+ h_{y,i}(\Omega)^2+h_{y,j}(\Omega')^2\Big).
\end{align*}
With these functions, we can find a form for average of energy density by expanding the exponential in Eq.\ref{qb2}. In this way, I find:
\begin{equation}\label{qb3}
\begin{split}
\la E\ra_{ev} &= \int_{\Omega}\int_{\Omega'}\la E(\Omega,\Omega')\ra
\\&=\frac{3A^2 \sum_{i,j=1}^{N_{\alpha}}1_{ij}}{4\pi r_L^2(1+3\xi)}\Big[1+\sum_{n=1}\frac{(-\eta)^n b_n}{(1+3\xi)^n}\Big]
\\&=\frac{3A^2 N_{\alpha}^2}{4\pi r_L^2(1+3\xi)}\Big[1+\sum_{n=1}\frac{(-\eta)^n b_n}{(1+3\xi)^n}\Big],
\end{split}
\end{equation}
where the coefficients $b_n$ are given by $b_1=1/8$, $b_2=1/20$, and so on. Additionally, Eq.\ref{qb3} allows us to establish the relationship between two forms of 3pF configurations$-$ spherical shape and cluster structures$-$ by comparing with Eq.\ref{q25}. Thus, we find the half-density radius $R_s$ as follows:
\begin{equation}
R_s =  r_L\sqrt{\frac{1+3\xi}{\mathcal{P}(\eta,\xi)}-3\xi},
\end{equation}  
where we have $A_s=A N_{\alpha}$ and $\mathcal{P}(\eta,\xi)=1+\sum_{n=1}\frac{(-\eta)^n b_n}{(1+3\xi)^n}$. We can see that this relation not only depends on $\eta$ but also on the size of cluster $r_L$. This means that we cannot determine the $R_s$ only having the ratio of $r_L/\ell_c$. Moreover, in this way, one can obtain the second moment of total energy density as follows:
\begin{equation}\label{qb5}
\begin{split}
\la E(\Omega,\Omega')^2 \ra&=\frac{3 A^2}{16 \pi ^2 r_L^4 \left(2+3\xi\right)\xi}\sum_{i,j}^{N_{\alpha}}e^{\frac{3\eta(\Delta_{1,1}^{ij}+\xi\Delta_{1,2}^{ij})}{24(2+3\xi)}}\\
&+\frac{3(A-1) A^2}{4\pi^2 r_L^4(1+8\xi+12\xi^2)}\sum_{i,j}^{N_{\alpha}} e^{\frac{3\eta(\Delta_{2,1}^{ij}+\xi\Delta_{2,2}^{ij})}{36(1+2\xi)}}\\
&+\frac{3A^3}{4\pi^2 r_L^4(1+8\xi+12\xi^2)}\sum_{i,j\neq j'}^{N_{\alpha}}e^{\frac{3\eta(\Delta_{3,1}^{ijj'}+\xi\Delta_{3,2}^{ijj'}+\xi^2\Delta_{3,3}^{ijj'})}{36(1+8\xi+12\xi^2)}}\\
&+\frac{9 (A-1)^2 A^2}{16 \pi ^2 r_L^4 \left(1+3\xi\right)^2}\sum_{i,j}^{N_{\alpha}}e^{\frac{3\eta(\Delta_{4,1}^{ij}+\xi\Delta_{4,2}^{ij})}{12(1+3\xi)}}
\\&+\frac{9 A^3}{16 \pi ^2 r_L^4 \left(1+3\xi\right)^2}\sum_{i\neq  i',j\neq j'}^{N_{\alpha}}e^{\frac{3\eta(\Delta_{5,1}^{ii'jj'}+\xi\Delta_{5,2}^{ii'jj'})}{24(1+3\xi)}}
\\&+\frac{9 (A-1) A^3}{8 \pi ^2 r_L^4 \left(1+3\xi\right)^2}\sum_{i\neq i',j}^{N_{\alpha}}e^{\frac{3\eta(\Delta_{6,1}^{ii'j}+\xi\Delta_{6,2}^{ii'j})}{24(1+3\xi)}},
\end{split}
\end{equation}
where $\Delta$ are the functions of the Euler angles $\Omega$ and $\Omega'$, such that they can be expressed as combinations of $f$, $h_{x}$, and $h_{y}$ defined in Eq.\ref{fi}-\ref{hy}, similar to what I found for $\la E(\Omega,\Omega')\ra$. By integrating out the Euler angles $\Omega$ and $\Omega'$ from Eq.\ref{qb5}, we find:
\begin{equation}\label{qb6}
\begin{split}
\la E^2\ra_{ev}&=\int_{\Omega}\int_{\Omega'}\la E(\Omega,\Omega')^2 \ra
\\&=\frac{3 A^2 N_{\alpha}^2}{16 \pi ^2 r_L^4 \left(2+3\xi\right)\xi}\mathcal{P}_1(\eta,\xi)\\
&+\frac{3(A-1) A^2 N_{\alpha}^2}{4\pi^2 r_L^4(1+8\xi+12\xi^2)}\mathcal{P}_2(\eta,\xi)\\
&+\frac{3A^3 N_{\alpha}^2(N_{\alpha}-1)}{4\pi^2 r_L^4(1+8\xi+12\xi^2)}\mathcal{P}_3(\eta,\xi)\\
&+\frac{9 (A-1)^2 A^2 N_{\alpha}^2(N_{\alpha}-1)}{16 \pi ^2 r_L^4 \left(1+3\xi\right)^2}\mathcal{P}_4(\eta,\xi)
\\&+\frac{9 A^3 N_{\alpha}^2(N_{\alpha}-1)^2}{16 \pi ^2 r_L^4 \left(1+3\xi\right)^2}\mathcal{P}_5(\eta,\xi)
\\&+\frac{9 (A-1) A^3 N_{\alpha}^2}{8 \pi ^2 r_L^4 \left(1+3\xi\right)^2}\mathcal{P}_6(\eta,\xi),
\end{split}
\end{equation}
where the functions $\mathcal{P}_i(\eta,\xi)$ are given by:
\begin{align*}
\mathcal{P}_1(\eta,\xi)&=\Big[1+\sum_{n=1}\frac{(-\eta)^n d_n^{(1)}}{(2+3\xi)^n}\Big],\\
\mathcal{P}_2(\eta,\xi)&=\Big[1+\sum_{n=1}\frac{(-\eta)^n d_n^{(2)}}{(1+2\xi)^n}\Big],\\
\mathcal{P}_3(\eta,\xi)&=\Big[1+\sum_{n=1}\frac{(-\eta)^n( d_n^{(3,1)}+d_n^{(3,2)} \xi)^n}{(1+8\xi+12\xi^2)^n}\Big],\\
\mathcal{P}_4(\eta,\xi)&=\Big[1+\sum_{n=1}\frac{(-\eta)^n d_n^{(4)}}{(1+3\xi)^n}\Big],\\
\mathcal{P}_5(\eta,\xi)&=\Big[1+\sum_{n=1}\frac{(-\eta)^n d_n^{(5)}}{(1+3\xi)^n}\Big],\\
\mathcal{P}_6(\eta,\xi)&=\Big[1+\sum_{n=1}\frac{(-\eta)^n d_n^{(6)}}{(1+3\xi)^n}\Big].
\end{align*} 	
The coefficients $d_n^{(i)}$ can be found in the same manner as $b_n$ obtained in the above. Having Eqs.\ref{qb3} and \ref{qb6}, I find the variance of total system energy density: 
\begin{equation}
\begin{split}
\text{var}(E/\la E\ra)&=\frac{\la E^2\ra_{ev}}{\la E\ra_{ev}^2}-1\\&=\frac{1}{3 A^2 N_{\alpha}^2}\Big[
\frac{(1+3\xi)^2}{ \left(2+3\xi\right)\xi}\frac{\mathcal{P}_1(\eta,\xi)}{\mathcal{P}^2(\eta,\xi)}\\
&+\frac{4(A-1)(1+3\xi)^2 }{(1+8\xi+12\xi^2)}\frac{\mathcal{P}_2(\eta,\xi)}{\mathcal{P}^2(\eta,\xi)}\\
&+\frac{4A(N_{\alpha}-1)(1+3\xi)^2}{(1+8\xi+12\xi^2)}\frac{\mathcal{P}_3(\eta,\xi)}{\mathcal{P}^2(\eta,\xi)}\\
&+3 (A-1)^2 (N_{\alpha}-1)\frac{\mathcal{P}_4(\eta,\xi)}{\mathcal{P}^2(\eta,\xi)}
\\&+3 A (N_{\alpha}-1)^2\frac{\mathcal{P}_5(\eta,\xi)}{\mathcal{P}^2(\eta,\xi)}
\\&+6 (A-1) A \frac{\mathcal{P}_6(\eta,\xi)}{\mathcal{P}^2(\eta,\xi)}\Big]-1.
\end{split}
\end{equation}
As can be seen, we cannot find a well defined form for the variance, unlike the spherical case obtained in Eq.\ref{varS}. Additionally, one can easily obtain $\varepsilon_2\{2\}$ using this approach.

\section{Cumulant formulation of fireball energy-density fluctuations}\label{app3}
I introduce the generating functional for the initial energy density in the transverse plane as \cite{Gronqvist:2016hym}:
\begin{equation}
Z[j]\equiv \la \exp\Big(\int d^2\mathbf{r}\; j(\mathbf{r})\epsilon(\mathbf{r}) \Big)\ra,\quad\quad W[j]=\ln Z[j],
\end{equation} 
where $j(\mathbf{r})$ and $\epsilon(\mathbf{r})$ are the auxiliary source and total system energy density. Connected correlation functions (cumulants) are generated by functional derivatives of $W[j]$:      
\begin{equation}
k_n(\mathbf{r}_1,\cdots,\mathbf{r}_n)=\frac{\delta^n W[j]}{\delta j(\mathbf{r}_1)\cdots\delta j(\mathbf{r}_n)}\Big|_{j=0}.
\end{equation}
The variance of total deposited energy,
$$\la E\ra = \int d^2\mathbf{r}\; \epsilon(\mathbf{r}),$$
is given by the second cumulants,
\begin{equation}
\text{var}(E)=\int d^2\mathbf{r}_1 d^2\mathbf{r}_2\; k_2(\mathbf{r}_1,\mathbf{r}_2),
\end{equation}
where
\begin{equation}\label{k2eq} 
k_2(\mathbf{r}_1,\mathbf{r}_2)=\la\epsilon(\mathbf{r}_1)\epsilon(\mathbf{r}_2)\ra-\la\epsilon(\mathbf{r}_1)\ra\la\epsilon(\mathbf{r}_2)\ra.
\end{equation}
\textit{$-$Decomposition into one- and two-body contributions:} Following the microscopic construction used in the main text, the energy density can be written as a sum of a one-body (mean-field) part ($\epsilon_1(\mathbf{r})$) and a genuine two-body part ($\epsilon_2(\mathbf{r})$),
$$\epsilon(\mathbf{r})=\epsilon_1(\mathbf{r})+\epsilon_2(\mathbf{r}).$$
Inserting this into the definition of $k_2$ in Eq.\ref{k2eq} gives:
\begin{equation}
\begin{split}
k_2(\mathbf{r}_1,\mathbf{r}_2)=&\la(\epsilon_1(\mathbf{r}_1)+\epsilon_2(\mathbf{r}_1))(\epsilon_1(\mathbf{r}_2)+\epsilon_2(\mathbf{r}_2)) \ra\\&-\la \epsilon_1(\mathbf{r}_1)+\epsilon_2(\mathbf{r}_1)\ra\la \epsilon_1(\mathbf{r}_2)+\epsilon_2(\mathbf{r}_2)\ra.
\end{split} 
\end{equation}
The pure one-body contribution cancels, $\la \epsilon_1(\mathbf{r}_1) \epsilon_1(\mathbf{r}_2)\ra - \la \epsilon_1(\mathbf{r}_1)\ra\la \epsilon_1(\mathbf{r}_2)\ra=0$, so that only cross and two-body terms survive:
\begin{equation}
\begin{split}
k_2(\mathbf{r}_1,\mathbf{r}_2)&= \la \epsilon_1(\mathbf{r}_1)\epsilon_2(\mathbf{r}_2)\ra-\la \epsilon_1(\mathbf{r}_1)\ra\la\epsilon_2(\mathbf{r}_2)\ra
\\&+\la \epsilon_2(\mathbf{r}_1)\epsilon_1(\mathbf{r}_2)\ra-\la \epsilon_2(\mathbf{r}_1)\ra\la\epsilon_1(\mathbf{r}_2)\ra
\\&+\la \epsilon_2(\mathbf{r}_1)\epsilon_2(\mathbf{r}_2)\ra-\la \epsilon_1(\mathbf{r}_2)\ra\la\epsilon_2(\mathbf{r}_2)\ra.
\end{split}
\end{equation}
\textit{$-$Alpha cluster decomposition:} For $\alpha$-clustered nuclei the two-body part further split into contributions from nucleons belonging to the same $\alpha$ cluster and to different $\alpha$ clusters,
\begin{equation}
\epsilon_2= \epsilon_2^{(\text{same}\;\alpha)}+\epsilon_2^{(\text{diff}\;\alpha)}.
\end{equation}
The two-body cumulant therefore becomes:
\begin{equation}
k_2^{(2\times 2)}=k_2^{(\text{same}\times\text{same})}+k_2^{(\text{diff}\times\text{diff})}+2k_2^{(\text{same}\times\text{diff})},
\end{equation}
which corresponding precisely to the intra-$\alpha$, inter-$\alpha$, and mixed contributions of the two-body density discussed in Sec.\ref{corr}. Finally, the variance of the total fireball energy is:
\begin{equation}
\begin{split}
\text{var}(E)=&\int d^2\mathbf{r}_1 d^2\mathbf{r}_2\; \\&\times\Big[k_2^{(1\times 2)}+k_2^{(\text{same}\times\text{same})}+k_2^{(\text{diff}\times\text{diff})}+2k_2^{(\text{same}\times\text{diff})}\Big],
\end{split}
\end{equation}
showing explicitly how $\alpha$-cluster substructure enters the fireball fluctuations through both internak cluster correlations and the geometry of cluster-cluster correlations. 
\bibliography{hydr.bib}
		
\end{document}